\documentclass[12pt]{article}

\usepackage{graphicx}
\usepackage{enumerate}
\usepackage[authoryear]{natbib}
\usepackage{url} 
\usepackage{amsfonts,amsmath,amsbsy,amsthm,amssymb,mathrsfs}
\usepackage{cases} 
\usepackage{anysize,indentfirst,geometry,xcolor}
\usepackage{authblk}

\usepackage{bm}
\usepackage{bigints} 
\usepackage{tikz}
\usepackage{caption}
\usepackage{subcaption}
\usepackage{multicol} 
\usepackage{multirow}
\usepackage[figuresright]{rotating}

\usepackage{setspace}

\theoremstyle{plain}
\newtheorem{prop}{Proposition}
\newtheorem{assum}{Assumption}

\def\bfZ{\mathbf{Z}}

\def\bfbeta{\bm{\beta}}

\def\hatalpha{\widehat{\alpha}}

\def\Pr{\operatorname{Pr}}

\title{Exploring causal effects of hormone- and radio-treatments in an observational study of breast cancer using copula-based semi-competing risks models}

\author{Tonghui Yu$^1$, Mengjiao Peng$^2$, Yifan Cui$^3$ , Elynn Chen$^{4}$, Chixiang Chen$^{5,6}$\thanks{chixiang.chen@som.umaryland.edu}, 
\hspace{.2cm} \\ 
$^1$ School of Physical and Mathematical Sciences, Nanyang Technological University, Singapore.\\
$^2$ School of Statistics, East China Normal University, China.\\
$^3$ Center for Data Science, Zhejiang University, China.\\
$^4$ Stern School of Business, New York University, New York, NY, USA.\\
$^5$ Division of Biostatistics and Bioinformatics, Department of Epidemiology and Public Health, University of Maryland School of Medicine, Baltimore, MD, U.S.A.\\
$^6$ Department of Neurosurgery, University of Maryland School of Medicine, Baltimore, MD, USA.
}

\date{}

\begin{document}
\maketitle

\begin{abstract}
Breast cancer patients after surgery may suffer from relapse or death in the duration of follow-up. Complexity arises in the analysis when the relapse faces potential dependent censoring. This phenomenon, known as semi-competing risk, demands advanced statistical tools for unbiased analysis. Despite progress in estimation and inference within semi-competing risks regression, its application to causal inference is still in its early stages. This article aims to establish a frequentist and semi-parametric framework based on copula models, facilitating valid causal inference, net quantity estimation, and sensitivity analysis for unmeasured factors under right-censored semi-competing risks data. We also study the non-parametric identification and propose novel procedures to enhance parameter estimation. We apply the proposed framework to a breast cancer study and detect the time-varying causal effects of hormone- and radio-treatments on patients' relapse-free survival and overall survival. Extensive numerical evaluations demonstrate the method's feasibility, highlighting minimal estimation bias and reliable statistical inference. 

\end{abstract}


\section{Introduction}

\label{s:intro}
\subsection{Semi-competing risks data}
 In clinical and biomedical studies, time-to-event endpoints often play a pivotal role, offering a tangible measure of the impact and effectiveness of medical interventions over time \citep{cohen2022clinical}. 
However, the analysis of time-to-event endpoints becomes intricate in the presence of semi-competing risks, a scenario frequently observed in various observational/clinical studies \citep{mao2021statistical}. This paper focuses on studying the effects of hormone and radiotherapies in the treatment of breast cancer patients. While these two therapies are widely adopted in clinical practice to improve survival for patients with breast cancer, it remains unclear whether the combination of hormone and radiotherapies is more beneficial to patients after surgery than using hormone therapy alone \citep{mayer2014,mcduff2022optimizing}. To unbiasedly address this question while accounting for different time-to-event endpoints during breast cancer progression, we aim to investigate both relapse-free survival (RFS, i.e., the time from the initial diagnosis to the time of tumor loco-regional or distant relapse) and overall survival (OS) and develop a valid causal inference tool to assess the effects of hormone and radiotherapies on these semi-competing risks.

Formally, semi-competing risks survival times encompass both the time to a non-terminal event (e.g., RFS) and the time to a terminal event (e.g., OS) from the study entry, denoted by $T_1$ and $T_2$, respectively. The pair $(T_{1},T_{2})$ can be conceptualized as a bivariate failure time, where $T_{2}$ acts as a censoring variable for $T_{1}$, but the reverse is not true—$T_{1}$ does not censor $T_{2}$. 
Unlike the classic competing risk setting \citep{fine1999competing,young2020} focusing on the first event and ignoring the information after the nonterminal event, the two endpoints in the semi-competing risk setting focus on both outcomes in a sequential manner and interplay between these outcomes \citep{varadhan2014semicompeting}. In our application, we are particularly interested in the time to cancer relapse and time to death, both of which are highly valued by patients and their caregivers. 
Understanding how factors influence both events and how two events interact would inform clinical decisions and promote dynamic predictions on the terminal event \citep{li2016flexible}. The above motivation underscores the necessity for specialized statistical methods that appropriately account for the interplay between these two distinct time-to-event processes in the analysis of semi-competing risks.


\subsection{Semi-parametric regression with semi-competing risks}
Semi-parametric regressions within the framework of semi-competing risks are of importance and can be divided into two distinct categories: crude quantities and net quantities \citep{varadhan2014semicompeting}. 
Concerning crude quantities, the sub-distribution approach and multi-state approach (specifically the so-called illness-death approach) are two kinds of typical means to handle survival data subject to dependently censored data. The sub-distribution approach \citep{fine1999competing,li2014subdistribution} evaluates the probability of each event occurring as the first one. Accordingly, it treats semi-competing risks akin to conventional competing risks. The illness-death approach \citep{xu2010illnessdeath}  encompasses two cause-specific hazard models, quantifying the impact of covariates on the occurrence of either the terminal or non-terminal event, along with a Markov model that characterizes the progression from the non-terminal to the terminal event.
However, both approaches focus on observable quantities rather than the marginal distributions of specific event times.
According to the bounds established by \cite{peterson1976bound}, the marginal distribution of $T_1$ is bounded from below by the sub-distribution function of the non-terminal event and from above by the distribution of the first event occurring. Such wide bounds suggest that predicting the non-terminal event time based solely on cause-specific hazards in the illness-death model could lead to biases in the case of heavily dependent censoring. Moreover, such shared frailty models encounter challenges in effectively describing the negative correlation between the two event times \citep{varadhan2014semicompeting}.

The net-quantity-based approaches concern primarily the ``net" effects of potential covariates on the marginal distribution of non-terminal and terminal event times. The marginal distribution in the absence of external influences, that is, hypothesizing the removal of the terminal event, may hold more direct clinical implications \citep{varadhan2014semicompeting}. Marginal methodologies include joint models specified through copulas \citep{peng2007semicompeting,chen2012semicompeting,peng2018semicompeting} or two separate marginal models augmented with artificial censoring techniques \citep{peng2006artificial,Ding2009artificial}. 
The latter approaches concentrate on estimating regression coefficients in marginal models of $T_1$ and $T_2$.
They, however, fail to provide insights into the dependence between the non-terminal and terminal events. 
Additionally, the marginal distribution of the non-terminal event time faces an issue of unidentifiability without making substantial additional assumptions.  
Therefore, copula-based models may offer a more meaningful framework, given their capacity to simultaneously explore marginal distributions and quantify the association between the two event times \citep{emura2021conditional,sun2023penalised,wei2023bivariate,yu2023unified}.

\subsection{Causal inference under semi-competing risks}

Several efforts have been made for causal interpretation under survival data with competing risks in literature \citep{young2020,stensrud2021generalized,stensrud2022}. 
However, the application of semi-competing risks to causal inference is still in its early stages. A notable challenge arises when comparing outcomes between treated and untreated groups after treatment initiation, where the studied subjects may experience death before reaching the target event, and the death rates in the treated and control groups may provide informative insights. Consequently, the conventional average causal effect becomes neither well-defined nor easily interpretable in a causal context with the exception that under additional assumptions  \citep{Nevo2022semicompeting, axelrod2022sensitivity}. \cite{robins1995analytic} showed the identifiability of average causal effect when censoring time is always observed, and without distinguishing between censoring due to death or other reasons. 

Under the semi-competing risks setting, the Survivor Average Causal Effect (SACE) has gained popularity as an alternative and valid estimand, quantifying the causal effect within the stratum of individuals who would have survived under both treatment values \citep{frangakis2002principal}. \cite{comment2019survivor} and \cite{xu2022bayesian} extended SACE to semi-competing risk setup and estimated time-varying SACE via Bayesian methods. \cite{Nevo2022semicompeting} further investigated its nonparametric partial identifiability and proposed sensitivity analysis using frailty-based illness-death models. In addition to SACE, \cite{huang2021causal} cast the semi-competing risks problem in the framework of causal mediation modeling. Later, \cite{deng2023separable} and \cite{breum2024} studied the separable effects on transition rates between states in multi-state models. Following \cite{huang2021causal}'s idea, \cite{deng2024direct} defined a counterfactual cumulative incidence of the terminal event based on counting process techniques. In this article, we provide an alternative approach to investigate treatment effects on either non-terminal or terminal event times.

Despite advancements, the feasibility and utility of integrating copula-based models into causal inference with semi-competing risks, along with its sensitivity analysis, remain unclear. As outlined earlier, copula-based models offer advantages in estimating net quantities and contribute to clinically meaningful interpretations. Herein, this article aims to establish a frequentist framework based on copula models to analyze our breast cancer data, aiming to facilitate valid causal interpretation, estimate net quantities, and conduct sensitivity analysis for semi-competing risks data.  







The rest of this article is organized as follows. Section \ref{sec:copula model} introduces basic notations for semi-competing risks data. Section \ref{sec:causal inference} introduces the causal framework and provides sensitivity analysis for unmeasured factors. Section \ref{section:estimation} summarizes estimation methods for regression parameters. Section \ref{sec:application} describes the analysis for breast cancer data. Section \ref{sec:simulation} displays extensive simulation studies to illustrate the proposed causal framework.
Section \ref{sec:conclusion} provides discussions and future extensions.

\section{Copula-based models for bivariate survival function}\label{sec:copula model}

In line with Sklar’s theorem \citep{sklar1959}, we establish a functional relationship between the bivariate survival function of $(T_1, T_2)$ and their respective marginal distributions as follows:
\begin{equation}
\Pr (T_1\geq t_1, T_2\geq t_2| \bfZ) = C\{S_1(t_1|\bfZ),S_2(t_2|\bfZ);\alpha\}, 0\leq t_1\leq t_2,
\label{model:jointdist}
\end{equation} 
where $S_k(t|\bfZ)$ is the survival functions of $T_k$ conditional on the covariate vector $\bfZ$, for $k=1,2$,  and $C\{u,v;\alpha\}$ denotes a specified copula function modeling inter-dependence between two conditional survival distributions with an unknown association parameter $\alpha$. This copula-based joint distribution is defined within the upper wedge where $T_1\leq T_2$. In contrast to models simply based on observable quantities, i.e., whichever occurs first, these marginal models of non-terminal and terminal events, whose details will be described in later sections, can evaluate ``net" covariate effects, purified effects on the non-terminal event $T_1$ with removed effects from $T_2$.  The  ``net" covariate effects hold notable clinical implications from the causal perspective in biomedical studies \citep{Ding2009artificial,chen2010}. 
In this article, we consider the Archimedean copula which can be expressed as
\begin{equation}
    C\{u,v;\alpha\} = \phi_{\alpha}^{-1}\{\phi_{\alpha}(u)+\phi_{\alpha}(v)\}, 0\leq u,v\leq 1,
    \label{model:arcm}
\end{equation}
where the generator function $\phi_{\alpha}: [0,1]\rightarrow [0,\infty)$ is continuous and strictly decreasing from $\phi_{\alpha}(0)>0$ to $\phi_{\alpha}(1)=0$. 
Consequently, Kendall’s $\tau$, a widely recognized metric for assessing the association between the two survival times, can be expressed as \citep{lakhal2008}
\begin{equation}
    \tau = 4\int_0^1 \phi_{\alpha}(u)/\phi'_{\alpha}(u) du+1.
\label{eq:kendal_tau}
\end{equation}
Upon the above models \eqref{model:jointdist}-\eqref{eq:kendal_tau}, we consider a sample of $n$ individuals. Let $\{T_{i1},T_{i2},C_i,\bfZ_i\}$, be $n$ independent copies of $\{T_1,T_2,C,\bfZ\}$, where $C_i$ is censoring time such that $T_{i1}$ and $T_{i2}$ are observed as $(X_{i},Y_{i},\delta_{i1},\delta_{i2})$. Here, $X_i =\min( T_{i1}, T_{i2},C_i)$ and $Y_i=\min(T_{i2}, C_i)$, 
$\delta_{i1}=I(X_i=T_{i1})$ 
and $\delta_{i2}= I(Y_i=T_{i2})$. 

The implementation of copula-based estimation techniques poses challenges due to their intricate nature. These approaches are often perceived as sophisticated and not extensively explored, particularly in the context of semi-competing risk causal inference. The remainder of this article aims to fill the gap and investigate the feasibility and potential advantages of applying the net-quantity-based semi-parametric framework to the field of causal inference. We begin by delving into the causal parameter specific to the context of semi-competing risks data. The estimation of regression parameters is elaborated upon in Section \ref{section:estimation}.

\section{Treatment effects}
\label{sec:causal inference}
\subsection{Observed data and Potential outcomes}

In clinical and biomedical research, investigators are often interested in evaluating the causal impact of treatments on both slowing the advancement of disease and extending overall survival following an intermediate event. To address these questions, we employ the potential outcomes framework \citep{neyman1923applications,rubin1974} to define causal effects. The data is then distinguished as observed data and potential outcomes.

\textbf{Observed data.} 
In addition to $(T_{i1}, T_{i2})$ and their observations $(X_{i}, Y_{i},\delta_{i1},\delta_{i2})$ defined in the previous section, there exists a treatment variable $A_i$ indicating whether individual $i$ received treatment or not. It is assumed that $\bfZ_i$ are pre-treatment covariates unaffected by the treatment assignment. Consequently, the observed data is denoted as $\mathcal{O}_i = \{X_{i},Y_{i},\delta_{i1},\delta_{i2}, A_i,\bfZ_i \}$, $i=1,\cdots,n$.


\textbf{Potential outcomes.} Before defining the potential outcomes, we first take into account the stable unit treatment value assumption (SUTVA) \citep{rubin1990comment}:
\begin{assum}
There is no interference between patients, and each treatment value corresponds to a single and consistent outcome, without the existence of multiple variations.
\label{assum:sutva}
\end{assum} 
Let $(T_{i1}(a),T_{i2}(a))$ be the potential time-to-event outcomes when the individual $i$ is subjected to treatment $A=a$ and $a=0$ or $1$. This establishes pairs of potential outcomes  $\{(T_{i1}(0),T_{i2}(0)),(T_{i1}(1),T_{i2}(1))\}$ for each individual. 
The causal effect pertains to the comparison between potential outcomes for either $T_1$ or $T_2$ within the same group of subjects, each under the influence of two competing treatments. We notice here that, within the framework of causal inference, at most one of the potential outcomes can be observed for each individual, while the others remain unobserved.  The SUTVA suggests that the observed event times $T_1$ and $T_2$ can be represented as linear combinations of the potential outcomes under different treatments, specifically $T_1=T_1(1)A+T_1(0)(1-A)$ and $T_2 = T_2(1)A+T_2(0)(1-A)$. 
Similarly, in the presence of censoring, denoted by $C$, the underlying potential outcomes $(T_{i1}(a),T_{i2}(a))$ could be observed in the form of $(X_{i}(a),Y_{i}(a))$ accompanied by the corresponding censoring indicators $(\delta_{i1}(a),\delta_{i2}(a))$. 
For further defining causal estimands, we adopt the no-unmeasured confounding assumption, which is often imposed in literature: 
\begin{assum} 
(i) (Conditional randomization) $A\perp (T_1(a), T_2(a), C)|\bfZ$
and (ii) (Non-informative censoring)  $C\perp (T_1(a), T_2(a))|\bfZ$.
\label{assum:random}
\end{assum}

We notice here that the Assumption \ref{assum:random} (ii) is less restrictive as it initially appears. Given the fact that the nonterminal event time is subject to either terminal event time or random censoring $C$, as a result, informative censoring is applied to the nonterminal event, while noninformative censoring is applied to the terminal event.

\subsection{Causal estimands}
An appropriate causal estimand should involve a valid causal interpretation, ensuring comparable individual profiles between treated and control groups throughout the time course \citep{Nevo2022semicompeting}. One popular approach to achieving this is by employing the concept of principal stratification \citep{frangakis2002principal,rubin2006causal,ding2011identifiability,dai2012partially}. Specifically, to account for censoring, the stratum-specific survivor causal effect (SCE) of treatment on the non-terminal event time is defined by the comparison of potential outcomes of $T_1$ among individuals who would always experience the non-terminal event before death regardless of treatment exposure \citep{tchetgen2014identification}. To account for dependent censoring, as described by \cite{Nevo2022semicompeting}, individuals with potential outcomes having $T_1(0)\leq T_2(0)$ and $T_1(1)\leq T_2(1)$ fall into the ``always-diseased" (AD) stratum. If a unit has $T_1(a)\geq T_2(a)$ for both $a=0,1$, it suggests that he/she is never-diseased (ND). 
Given the considerable interest in the survival probability of non-terminal events in clinical studies, we primarily focus on the following causal estimand:
\begin{equation}
\begin{split}
\operatorname{AD-SCE}_1(t)
&=\Pr(T_1(1)\geq t|T_1(0)\leq T_2(0),T_1(1)\leq T_2(1))\\
&-\Pr(T_1(0)\geq t|T_1(0)\leq T_2(0),T_1(1)\leq T_2(1)).
\label{eq:SCE1}
\end{split}
\end{equation}
Herein, we study the time-varying SCE estimands in Eq \eqref{eq:SCE1} rather than its integration version (or restricted survival mean) defined in \cite{Nevo2022semicompeting} because estimators of survival means may lead to inevitably inaccurate and unsatisfactory results caused by point predictions in many finite-sample numerical examples \citep{graf1999assessment}. The goal of this article is to explore varying  (dynamic) treatment effects over time. In addition, 
since the terminal event could occur after or before the non-terminal event, 
we consider two types of stratum-specific SCE for the terminal event:
\begin{equation}
\begin{split}
\operatorname{AD-SCE}_2(t)
&=\Pr(T_2(1)\geq t|T_1(0)\leq T_2(0),T_1(1)\leq T_2(1))\\
&-\Pr(T_2(0)\geq t|T_1(0)\leq T_2(0),T_1(1)\leq T_2(1)),
\label{eq:SCE2}
\end{split}
\end{equation}

\begin{equation}
\begin{split}
\operatorname{ND-SCE}_2(t)
&=\Pr(T_2(1)\geq t|T_1(0)\geq T_2(0),T_1(1)\geq T_2(1))\\
&-\Pr(T_2(0)\geq t|T_1(0)\geq T_2(0),T_1(1)\geq T_2(1)).
\label{eq:SCE3}
\end{split}
\end{equation}

We remark here that the estimands above are subject to the joint distribution of event times from both worlds, i.e., $a=0,1$, which cannot be observed simultaneously. To facilitate non-parametric identification, we introduce a combined covariate vector $\bfZ$ to describe the alterations in the occurrences of both event times, and impose the following assumption. 
\begin{assum}
The cross-world dependence is captured by pre-treatment covariates $\bfZ$ in the sense that $(T_1(0),T_2(0))\perp (T_1(1),T_2(1))|\bfZ$.
\label{assum: covariate}
\end{assum}


We notice here that the cross-world conditional independence in Assumption \ref{assum: covariate} is strong and untestable in practice \citep{richardson2013single,robins2022interventionist,stensrud2021discussion}.
More discussion about cross-world unmeasured factors and a weaker assumption can be found in Section \ref{section:sensitivity}. Under Assumptions \ref{assum:sutva}-\ref{assum: covariate}, we establish the identification of causal estimands of our interest in Proposition 1, with proof provided in Supplementary Section 1. The above estimands \eqref{eq:SCE1}-\eqref{eq:SCE3} can alternatively be defined across observable strata, such as $\{X(0) \leq Y(0), X(1) \leq Y(1)\}$ instead of the ``AD" stratum, and $\{X(0) \geq Y(0), X(1) \geq Y(1)\}$ instead of the ``ND" stratum, and it can be simply verified that these two types of definitions are equivalent under noninformative censoring in Assumption \ref{assum:random}.

\begin{prop}
\label{prop:scae}
(Non-parametric identification) Under Assumptions \ref{assum:sutva}-\ref{assum: covariate}, the stratum-specific survivor average causal effects in \eqref{eq:SCE1}-\eqref{eq:SCE3} are identified by
\begin{equation}
\operatorname{AD-SCE}_k(t)
= \frac{\mathbb{E}_{\bfZ}\left[(S_{k|T_1\leq T_2,A=1,\bfZ}(t)-S_{k|T_1\leq T_2,A=0,\bfZ}(t) )\Pi_{A=0,\bfZ}\Pi_{A=1,\bfZ} \right]}{\mathbb{E}_{\bfZ}\left[\Pi_{A=0,\bfZ}\Pi_{A=1,\bfZ} \right]}, k=1,2,
\label{eq:adSCE}
\end{equation}
and 
\begin{equation}
\operatorname{ND-SCE}_2(t)
= \frac{\mathbb{E}_{\bfZ}\left[ (S_{2|T_1\geq T_2,A=1,\bfZ}(t)-S_{2|T_1\geq T_2,A=0,\bfZ}(t))\widetilde{\Pi}_{A=0,\bfZ}\widetilde{\Pi}_{A=1,\bfZ} \right]}{\mathbb{E}_{\bfZ}\left[\widetilde{\Pi}_{A=0,\bfZ}\widetilde{\Pi}_{A=1,\bfZ} \right]}, t>0,
\label{eq:ndSCE}
\end{equation}
where $S_{k|\mathcal{Q}}(t)=\Pr(T_k\geq t|\mathcal{Q})$, $\Pi(\mathcal{Q}) = \Pr(T_1\leq T_2|\mathcal{Q})$, $\widetilde{\Pi}(\mathcal{Q}) = \Pr(T_1\geq T_2|\mathcal{Q})$ for any event $\mathcal{Q}$.
\end{prop}

The above proposition justifies the validity and identifiability of the proposed estimands. Before concluding this section, we provide a brief illustration of applying copula techniques to identify these estimands. A detailed estimation procedure is outlined in Section \ref{section:estimation}. To facilitate, we require the following assumption:
\begin{assum}
Conditional on covariates $\bfZ$, the joint distribution function of $(T_1(a),T_2(a))$ follows a copula model
\begin{equation}
D(t_1,t_2|a,\bfZ) = \Pr (T_1\geq t_1, T_2\geq t_2| A=a, \bfZ)
= C^{(a)}\{S_1(t_1|a,\bfZ),S_2(t_2|a,\bfZ);\alpha^{(a)}\}, \quad 
\label{model: copula}
\end{equation}
where $0\leq t_1\leq t_2$, $S_k(t|a,\bfZ)$ is the survival functions of $T_k(a)$ ($k=1,2$) given treatment and covariates $\bfZ$; $C^{(a)}\{u,v;\alpha\}$ is a pre-specified copula function with $\alpha^{(a)}$ being a constant parameter describing extra association between event times under the treatment status $a$.
\label{assum:cross-world}
\end{assum}

It is noteworthy that Assumption \ref{assum:cross-world} permits the utilization of various copula structures corresponding to different treatment assignments. Consequently, the proposed model exhibits greater flexibility compared to parametric frailty models found in the existing literature \citep{Nevo2022semicompeting}. A weaker substitution of Assumption \ref{assum:cross-world} will be discussed in Section \ref{section:sensitivity}.

We will now present the identification formulas within the framework of copulas. To ease presentation, we define $C_1^{(a)}(u,v,\alpha) = \partial C^{(a)}(u,v;\alpha)/\partial u$, $C_2^{(a)}(u,v,\alpha) = \partial C^{(a)}(u,v;\alpha)/\partial v$,
$ D\{s,t|a,\bfZ\}=C^{(a)}\{S_1(s|a,\bfZ), S_2(t|a,\bfZ);\alpha^{(a)}\},$ 
$ D_{k}\{s,t|a,\bfZ\}=C_k^{(a)}\{S_1(s|a,\bfZ), S_2(t|a,\bfZ);\alpha^{(a)}\}$, $k=1,2$.
Then, under Assumptions  \ref{assum:sutva}-\ref{assum:cross-world}, by simple derivation, the quantities in Proposition 1 can be re-written as 
\begin{equation}
S_{1|T_1\leq T_2,A=a,\bfZ}(t)
= \frac{\int_t^{\infty} D_1 \left(s,s|a,\bfZ  \right)d S_1(s| a,\bfZ )}{\int_0^{\infty} D_1 \left(s,s|a,\bfZ  \right)d S_1(s| a,\bfZ )},
\label{eq:s1}
\end{equation}
\begin{equation}
S_{2|T_1\leq T_2,A=a,\bfZ}(t)
 = \frac{S_2(t| a,\bfZ )+\int_t^{\infty}D_2 \left(s,s|a,\bfZ  \right)d S_2(s| a,\bfZ )}{ 1+\int_0^{\infty}D_2 \left(s,s|a,\bfZ  \right)d S_2(s| a,\bfZ )},
\label{eq:s2}
\end{equation}
\begin{equation}
S_{2|T_1> T_2,A=a,\bfZ}(t)
= \frac{\int_t^{\infty} D_2 \left(s,s|a,\bfZ  \right)dS_2(s|a,\bfZ)}{\int_0^{\infty} D_2 \left(s,s|a,\bfZ  \right)dS_2(s|a,\bfZ)}. 
\label{eq:s22}
\end{equation}
The denominators in \eqref{eq:s1}-\eqref{eq:s22} are the explicit expressions of $\Pi_{\mathcal{Q}}$ and $\widetilde{\Pi}_{\mathcal{Q}}$ for event $\mathcal{Q}$. 
Quantities \eqref{eq:s1}-\eqref{eq:s22} use observable data in the whole upper wedge of $(T_1, T_2)$ and thus can be estimable. Plugging estimates of $\alpha^{(a)}$,$S_1(\cdot|a,\bfZ)$, and $S_2(\cdot|a,\bfZ)$ assisted  by methods in \cite{peng2007semicompeting} and \cite{zhu2022nee}, estimates of causal effects \eqref{eq:SCE1}-\eqref{eq:SCE3} can be obtained from \eqref{eq:adSCE}-\eqref{eq:ndSCE},\eqref{eq:s1}-\eqref{eq:s22}. We refer readers to Section \ref{section:estimation} for estimation details.

\subsection{Sensitivity analysis}\label{section:sensitivity}
 
Before describing the proposed estimation, this section directs the reader's focus to Assumptions \ref{assum: covariate}-\ref{assum:cross-world}: cross-world dependence is fully captured by the observed covariates $\bfZ$. While these assumptions may be valid in many applications, they may not hold and might not even be testable in a general context since cross-world data cannot be simultaneously observed for each patient. 
In practice, we recommend incorporating expert knowledge to guide the identification of crucial covariates while concurrently assessing the sensitivity of estimates to unmeasured factors \citep{axelrod2022sensitivity,Nevo2022semicompeting}. To this end, we propose the copula-based sensitivity analysis. To capture the remaining dependency attributed to unmeasured factors shared across two worlds, we introduce a frailty $\gamma$ and a weaker restriction than Assumptions \ref{assum: covariate}-\ref{assum:cross-world}. 
\begin{assum}
There exists an unmeasured time-unvarying frailty variable $\gamma>0$ such that the cross-world dependence is captured by pre-treatment covariates $\bfZ$ together with $\gamma$ in the sense that $(T_1(0),T_2(0))\perp (T_1(1),T_2(1))|(\gamma,\bfZ)$.
\label{assum:frailty}
\end{assum}

\begin{assum}
The time-unvarying frailty variable $\gamma>0$, from a distribution with mean 1 and variance $\sigma$, satisfies:
\begin{enumerate}[(i)]
\item The frailty variable operates multiplicatively on conditional cumulative hazard function of $T_k(a)$, that is, $\Lambda_k(t|A=a,\gamma,\bfZ) = \gamma\widetilde{\Lambda}_k(t|A=a,\bfZ)$,
where $\Lambda_k(t|A=a,\gamma,\bfZ) $ is the conditional cumulative hazard function of $T_k(a)$ given treatment $a$ and $(\gamma,\bfZ)$, and $\widetilde{\Lambda}_k(t|A=a,\bfZ)$ is irrelevant to $\gamma$.
\item The conditional joint distribution function of $(T_1(a),T_2(a))$, given $(\gamma,\bfZ)$, follows a copula model 
$D(t_1,t_2|a,\gamma,\bfZ) 
= C^{(a)}\{S_1(t_1|a,\gamma,\bfZ),S_2(t_2|a,\gamma,\bfZ);\alpha^{(a)}\}$ 
where $S_k(t|a,\gamma,\bfZ)=\exp[-\Lambda_k(t|a,\gamma,\bfZ)]$ is the survival functions of $T_k(a)$ ($k=1,2$) given treatment and $(\gamma,\bfZ)$, the definitions of $\alpha^{(a)}$ and $C^{(a)}\{u,v;\alpha\}$ are provided in Assumption \ref{assum:cross-world}.
\end{enumerate}
\label{assum:cross-world_frailty}
\end{assum}

The frailty variable in Assumption \ref{assum:frailty} is commonly used in the survival analysis literature to explain unobserved heterogeneity due to unmeasured covariates or other sources of natural variation.  It is also widely applied in many causal inference literature to account for the cross-world dependence \citep{aalen2015does,martinussen2020subtleties,axelrod2022sensitivity}, a plausible remedy when the cross-world conditional independence in Assumption \ref{assum: covariate} is violated. 
This assumption also implies their marginal hazard ratios at time t on treatment $A=1$ versus control $A=0$ for the same patient population with covariates $\bfZ$ and unmeasured factor $\gamma$ who would survive that time, no matter what treatment, which is related to the so-called conditional causal hazard ratio \citep{martinussen2020subtleties}.
We stress in Assumption \ref{assum:cross-world_frailty} (ii) that the unmeasured factor has impacts on marginal distributions except for the global association between two event times under each counterfactual arm. We refer readers to the discussion section for more general identification assumptions and possible extensions.

Based on the weaker assumptions \ref{assum:frailty}-\ref{assum:cross-world_frailty}, and coupled with  Assumptions \ref{assum:sutva}-\ref{assum:random}, the causal estimands in \eqref{eq:SCE1}-\eqref{eq:SCE3} are identifiable with detailed formulas in 
Supplementary Section 2.
It is important to note that the distribution of frailty $\gamma$ may not be estimable in general. As a result,  we evaluate the effect of unmeasured factors by pre-specifying values of the unidentifiable parameter $\sigma$. The expectation with respect to frailty as Eq.(A.2)-(A.4) in the Supplementary Material can be then numerically calculated by Monte Carlo integration techniques.


\section{Estimation}\label{section:estimation}
The estimation of copula-based semicompeting risks regression has been extensively explored by existing literature. 
Nevertheless, adapting this framework to address causal inference involves significant adjustments and remains inadequately studied, where the literature either exhibits bias in cases of heavily dependent censoring or encounters challenges in effectively portraying the negative correlation between the two event times \citep{peterson1976bound,varadhan2014semicompeting}.
Herein, we provide an improved procedure to estimate parameters in the copula-based causal model. 
To facilitate the estimation process,
we limit $C^{(a)}\{\cdot\}$ to the Archimedean copula class, and for illustration, we use (but not limited to) the proportional hazards models to marginally fit $T_1$ and $T_2$, respectively. 
\subsection{Estimation in semiparametric models without frailty}
\label{sec:estimation_no frailty}
In this subsection, we delve into the analysis of marginal semiparametric models without frailty, a prevalent approach in the literature on copula-based semi-competing risks models, with novel and enhanced estimation procedures.
Specifically,
we now turn to estimating parameter $\alpha^{(a)}$ and each marginal distributions by employing the following marginal models
\begin{equation}
\lambda_k(t|\bfZ,A=a) = \lambda_{0k}^{(a)}(t)\exp(\bfbeta_k^{(a)T}\bfZ),
\label{model: marginal1}
\end{equation}
where $\lambda_k(t|Z,A=a) $ is the conditional hazard function of $T_k(a)$, $k=1,2$, given the covariates $\bfZ$ and treatment $A=a$; $\bfbeta_k^{(a)}$ is a vector of unknown regression coefficients, and $\lambda_{0k}^{(a)}$ is an unspecified baseline hazard function. Let $ \Lambda_{0k}^{(a)}(t) = \int_0^t \lambda_{0k}^{(a)}(s)ds$ be the baseline cumulative hazard function. And the conditional survival functions are in the forms of $S_k(t|\bfZ,A)  =\exp[-\Lambda_{0k}^{(a)}(t)\exp(\bfbeta_k^{(a)T}\bfZ)]$. We denote $C_{21}(u,v,\alpha) =C_{12}(u,v,\alpha) = \partial^2 C(u,v;\alpha)/\partial u\partial v$ and
$ D_{jk}\{X_i,Y_i;\alpha^{(a)}\}=C_{jk}\{S_1(X_i|\bfZ_i,a,\gamma_i), S_2(Y_i|\bfZ_i,a,\gamma_i);\alpha^{(a)}\}$,   $j,k=1,2. $ 
Under models \eqref{model: copula} and \eqref{model: marginal1}, the log-likelihood function concerning the parameters to be estimated given observed data $\mathcal{O}=\{X_i, Y_i,\delta_{i1},\delta_{i2}, A_i,\bfZ_i:i=1,\cdots,n\}$ can be written as
\begin{equation*}
\begin{split}
&l(\alpha^{(a)},\Lambda_{01}^{(a)},\Lambda_{02}^{(a)},\bfbeta_{1}^{(a)},\bfbeta_{2}^{(a)})\\
&= \sum\limits_{i=1}^n  (1-\delta_{i1})(1-\delta_{i2})I(A_i=a)\log \left [ D\{X_i,Y_i;\alpha^{(A_i)}\}\right]+\delta_{i1}\delta_{i2}I(A_i=a)\log\left[ D_{12}\{X_i,Y_i;\alpha^{(A_i)}\}\right]\\
& +\delta_{i1}(1-\delta_{i2})I(A_i=a)\log \left [ D_{1}\{X_i,Y_i;\alpha^{(A_i)}\}\right]+ (1-\delta_{i1})\delta_{i2}I(A_i=a)\log \left [ D_{2}\{X_i,Y_i;\alpha^{(A_i)}\}\right]\\
&+\delta_{i1}I(A_i=a)\log\left[ S_1(X_i|\bfZ_i,A_i)\lambda_{01}^{(A_i)}(X_i)\exp(\bfbeta_1^{(A_i)T}\bfZ_i)  \right]\\
& + \delta_{i2}I(A_i=a)\log\left[S_2(Y_i|\bfZ_i,A_i)\lambda_{02}^{(A_i)}(Y_i)\exp(\bfbeta_2^{(A_i)T}\bfZ_i) \right].
\label{eq:cox_lik}
\end{split}
\end{equation*}
These parameters can be estimated by adapting the nonparametric maximum likelihood estimation (NPMLE) \citep{peng2018semicompeting,peng2019}. However, the objective function of the aforementioned NPMLE method
is only locally convex in a small neighborhood of the truth and thus sensitive to initial estimates. Poor initialization will lead to biases in the final estimation or even algorithm convergence issues. Accordingly, we improve the above NPMLE method 
by substituting initial values with estimates derived from \cite{peng2018semicompeting} and \cite{zhu2022nee}, which have proven effective in simulation studies. Such adaptation and integration can effectively reduce the number of iterations and computation time as well as estimation biases. 
It is worth noting that the proposed estimation is not limited to the time-independent Cox proportional hazards model aforementioned. We can easily extend it to other model forms, including the proportional odds model and varying-coefficient models \citep{peng2018semicompeting, zhu2022nee}. 

\subsection{Estimation with frailty (Sensitivity analysis)}
\label{sec:estimation_frailty}
Frailty terms are frequently introduced for sensitivity analysis or to encapsulate latent information, such as the remaining dependency attributed to unmeasured confounders shared across two worlds in semi-competing risks causal inference \citep{Nevo2022semicompeting}.
Consequently, we extend to study the estimation procedure under the setting described in Section \ref{section:sensitivity}. Note that incorporating a frailty, even with a pre-specified value of $\sigma$, 
 is non-trivial and necessitates more intricate estimation, thus warranting further methodological development. Suppose that the marginal regressions of $T_k(a)$ on $\bfZ$, $k = 1, 2$, follow the shared-frailty Cox proportional hazards models
\begin{equation}
\lambda_k(t|\bfZ,A=a, \gamma) = \gamma\lambda_{0k}^{(a)}(t)\exp(\bfbeta_k^{(a)T}\bfZ),
\label{model: marginal}
\end{equation}
where $\lambda_k(t|Z,A=a, \gamma) $ is the conditional hazard function of $T_k(a)$ given the covariates $\bfZ$ and treatment $a$; the definitions of $\bfbeta_k^{(a)}$ and $\lambda_{0k}^{(a)}$ are same as model \eqref{model: marginal1}; 
and $\gamma$ is the frailty defined in Assumption \ref{assum:cross-world_frailty}. 
The log-likelihood function regarding to the parameters $(\alpha^{(a)},\Lambda_{01}^{(a)},\Lambda_{02}^{(a)},\bfbeta_{1}^{(a)},\bfbeta_{2}^{(a)})$ given observed data $\mathcal{O}$ can be expressed by
\begin{equation}l_o(\alpha^{(a)},\Lambda_{01}^{(a)},\Lambda_{02}^{(a)},\bfbeta_{1}^{(a)},\bfbeta_{2}^{(a)};\sigma)
=\sum\limits_{i=1}^n \log\left\{ \int \exp[l_i(\alpha^{(a)},\Lambda_{01}^{(a)},\Lambda_{02}^{(a)},\bfbeta_{1}^{(a)},\bfbeta_{2}^{(a)})]f_{\gamma} (\gamma_i|\sigma)d\gamma_i\right\},
\label{eq:loglik}
\end{equation}
where 
$f_{\gamma}(\gamma|\sigma)$ is the density function of $\gamma$ with prespecified variance parameter $\sigma$, and  
\begin{equation*}
\begin{split}
&l_i(\alpha^{(a)},\Lambda_{01}^{(a)},\Lambda_{02}^{(a)},\bfbeta_{1}^{(a)},\bfbeta_{2}^{(a)})
= (1-\delta_{i1})(1-\delta_{i2})\log \left [ D\{X_i,Y_i;\alpha^{(A_i)}\}\right]\\
&+  \delta_{i1}\delta_{i2}\log\left[ D_{12}\{X_i,Y_i;\alpha^{(A_i)}\}S_1(X_i|\bfZ_i,A_i,\gamma_i)\lambda_{01}^{(A_i)}(X_i)\gamma_i\exp(\bfbeta_1^{(A_i)T}\bfZ_i)  \right]\\
& + \delta_{i1}\delta_{i2}\log\left[S_2(Y_i|\bfZ_i,A_i,\gamma_i)\lambda_{02}^{(A_i)}(Y_i)\gamma_i\exp(\bfbeta_2^{(A_i)T}\bfZ_i) \right]\\
& +\delta_{i1}(1-\delta_{i2})\log \left [ D_{1}\{X_i,Y_i;\alpha^{(A_i)}\}S_1(X_i|\bfZ_i,A_i,\gamma_i)\lambda_{01}^{(A_i)}(X_i)\gamma_i\exp(\bfbeta_1^{(A_i)T}\bfZ_i) \right]\\
& + (1-\delta_{i1})\delta_{i2}\log \left [ D_{2}\{X_i,Y_i;\alpha^{(A_i)}\}S_2(Y_i|\bfZ_i,A_i,\gamma_i)\lambda_{02}^{(A_i)}(Y_i)\gamma_i\exp(\bfbeta_2^{(A_i)T}\bfZ_i)\right].
\end{split}
\end{equation*}

The integral in \eqref{eq:loglik} poses analytical intractability, making a direct evaluation of the likelihood function challenging. A prevalent approach involves approximating it and subsequently utilizing the approximated likelihood to conduct inference about the model parameters.
We modify the MCEM algorithm proposed by \cite{peng2018semicompeting} to accommodate our causal framework. To reduce computation complexity, we start from a small MCMC sample size and increase it as computation progresses to approximate the likelihood above to derive estimates of $(\alpha^{(a)},\Lambda_{01}^{(a)},\Lambda_{02}^{(a)},\bfbeta_{1}^{(a)},\bfbeta_{2}^{(a)})$. We refer readers to Section 3 from the Supplementary Material for more implementation details. Unlike \cite{peng2018semicompeting}'s regression model, frailty variance is not estimable and should be prespecified in our potential outcomes framework given that the two-world outcomes cannot be simultaneously observed. Nevertheless, it is noteworthy that their theoretical results remain applicable to our setting. In addition, from  \eqref{eq:loglik}, the proposed method is based upon (but not limited to) gamma-frailty for illustration. The proposed method can be extended to other general survival models and frailty distributions as well.

\subsection{Statistical inference for causal parameters}
Utilizing the estimators derived from Sections \ref{sec:estimation_no frailty} and \ref{sec:estimation_frailty}, we proceed to estimate the causal effects via principal stratification in the context of semi-competing risks data.
If there is no unmeasured confounding, plugging estimators for $(\alpha^{(a)},\Lambda_{01}^{(a)},\Lambda_{02}^{(a)},\bfbeta_{1}^{(a)},\bfbeta_{2}^{(a)})$, $a=0,1$, obtained from Section \ref{sec:estimation_no frailty}, which are consistent and asymptotically normal \citep{peng2007semicompeting,zhu2022nee,peng2018semicompeting}, into \eqref{eq:s1}-\eqref{eq:s22} and replacing $S_{k|T_1\leq T_2, A=a,\bfZ}$ and $S_{2|T_1\geq T_2, A=a,\bfZ}$ in \eqref{eq:SCE1}-\eqref{eq:SCE3} with its empirical counterpart yields consistent and asymptotically normal estimators for the three SCEs including $\operatorname{ND-SCE}_2(t)$, $\operatorname{AD-SCE}_1(t)$, and $\operatorname{AD-SCE}_2(t)$ with each time point $t$. The asymptotic properties for the causal parameters follow from the continuous mapping theorem and the delta method. 
The same asymptotic properties are applied to the proposed method for sensitivity analysis.  
However, in analog to \cite{zhu2022nee} and \cite{peng2018semicompeting}, asymptotic variance of estimators of $(\alpha^{(a)},\Lambda_{01}^{(a)},\Lambda_{02}^{(a)},\bfbeta_{1}^{(a)},\bfbeta_{2}^{(a)})$ have no explicit expressions. Thus, the Bootstrapping method is more preferred in practice to compute the standard errors of regression and causal parameters.

\section{Application to a breast cancer study}
\label{sec:application}



 Hormone therapy or neoadjuvant endocrine therapy is often used in clinical practice for breast cancer patients, as an adjuvant treatment following surgery. Despite its effectiveness, there is considerable debate surrounding the decision of whether to combine it with chemotherapy or radiotherapy or to pursue separate therapy after surgery \citep{mayer2014}. Our objective was to apply our developed methods to evaluate the impact of hormone therapy and compare it with the combination of hormone therapy and radiotherapy in breast cancer patients who have undergone mastectomy surgery. 

\textbf{Data.} We applied our methods to analyze the data from the Molecular Taxonomy of Breast Cancer International Consortium (METABRIC) cohort \citep{curtis2016nature,pereira2016somatic}. 
The response variable of our interest was the pair of relapse-free survival (RFS, $T_1$) and overall survival (OS, $T_2$) outcomes measured in months (i.e., semi-competing risks). Due to loss to follow-up and dependent censoring, the overall censoring rates for RFS and OS were  55$\%$ and 34$\%$, respectively. 
 Covariates used in our analysis included age at diagnosis (21-96 years old), ER status (ER, 1=positive, 0=negative), PR status (PR, 1=positive, 0=negative), HER2 status (HER2,1=positive, 0=negative), inferred menopausal state (MENO, 1=post-menopausal, 0=pre-menopausal), whether the number of lymph nodes examined positive bigger than 0 or not (NODE, 1= yes, 0=no), Nottingham prognostic index (NPI), and tumor size (SIZE). After data processing, there were 1114 patients taking mastectomies with complete records for covariates, RFS, and OS. In the following analyses, we used the Frank copula to describe the joint distribution of RFS and OS.

\begin{figure}[h]
  \medskip
\includegraphics[height=16cm,width=18cm]{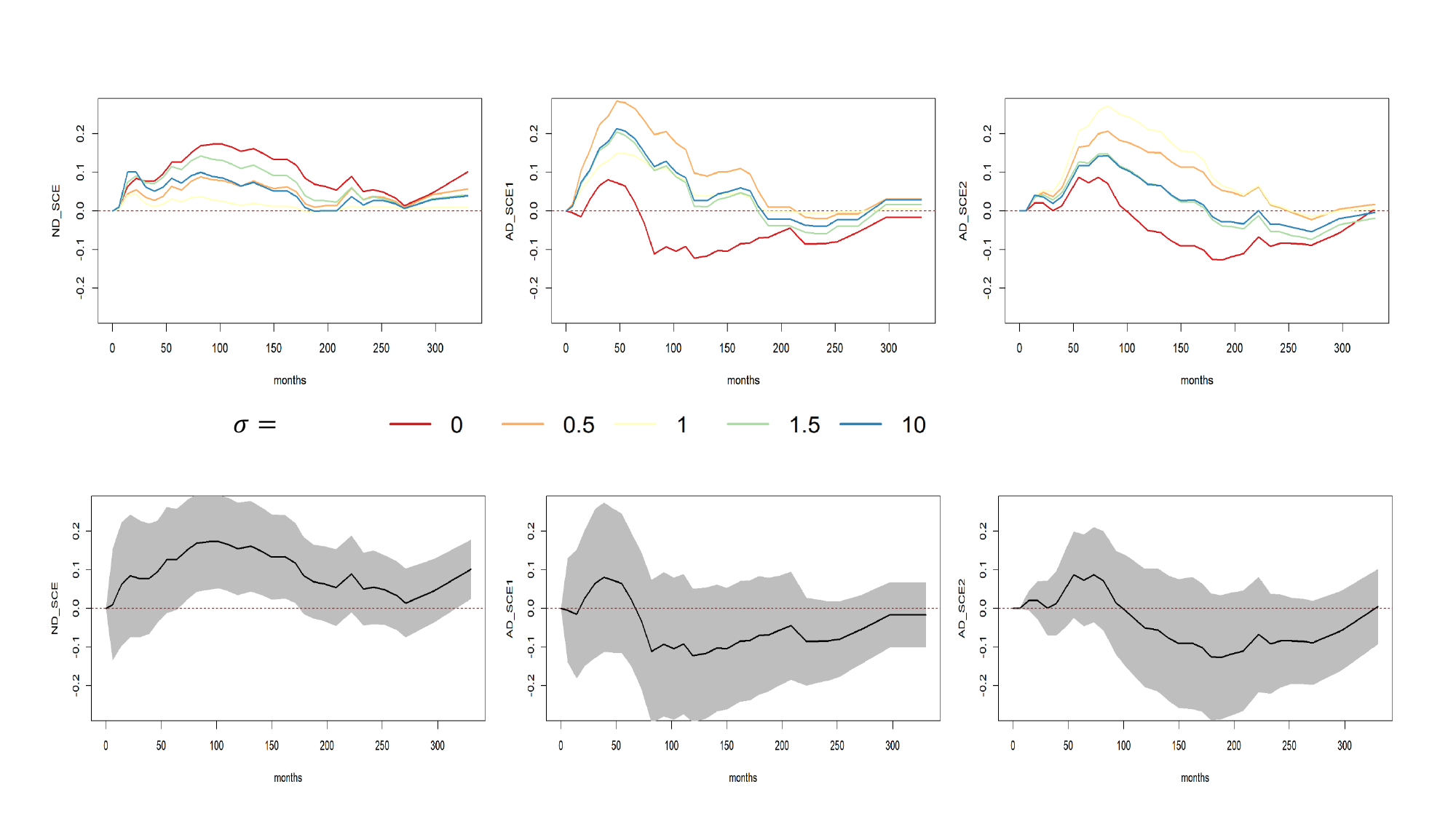}
\caption{Causal effects of hormone treatment alone compared to no treatment use in breast cancer. The three plots in the top row present the estimated $\operatorname{ND-SCE_2}$, $\operatorname{AD-SCE_1}$ and $\operatorname{AD-SCE_2}$ with different specified frailty variance $\sigma$ (sensitivity analysis). The three plots in the bottom row present the estimated causal parameters along with their 95$\%$ confidence interval under pre-specified $\sigma = 0$ (i.e., no unmeasured confounding).}
    \label{fig:causal_breast}
\end{figure}

\begin{figure}[h]
  \medskip
\includegraphics[height=16cm,width=18cm]{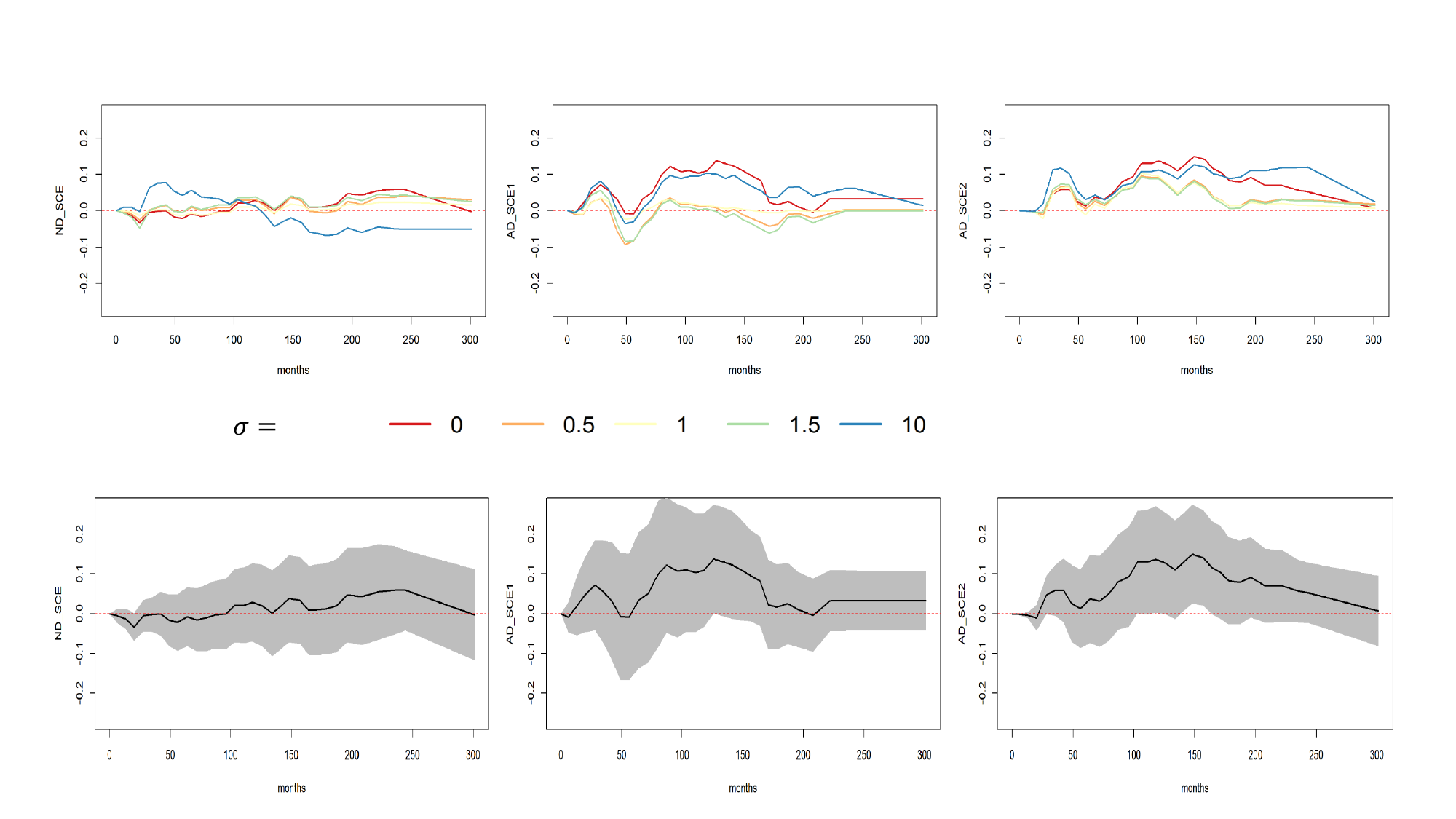}
\caption{Causal effects of hormone-and-radio treatment compared to hormone therapy alone in breast cancer. The three plots in the top row present the estimated $\operatorname{ND-SCE_2}$, $\operatorname{AD-SCE_1}$ and $\operatorname{AD-SCE_2}$ with different specified frailty variance $\sigma$ (sensitivity analysis). The three plots in the bottom row present the estimated causal parameters along with their 95$\%$ confidence interval under pre-specified $\sigma = 0$ (i.e., no unmeasured confounding).}
    \label{fig:causal_breast2}
\end{figure}

\textbf{Primary analysis.} We conducted two primary analyses: (1) We first quantified the causal effect of hormone treatment alone $(a=1)$ in comparison to no treatment $(a=0)$ using the proposed causal framework. To eliminate the influence and interaction of other treatments, we specifically focused on a subset of patients who did not receive chemotherapy or radiotherapy, leaving 611 patients ($59.4\%$ patients with hormone treatment alone vs $40.6\%$ without any treatment) in this analysis. 
The estimation of regression coefficients is summarized in Supplementary Table S.4. We dropped the variables PR and HER2 in this analysis, considering that the estimated coefficients for PR and HER2 status were very small in this limited sample.
Figure \ref{fig:causal_breast} displays the curves of causal parameters varying with time under two principle strata (AD, ND). 
(2) We continued to study the causal effect of the combination of hormone and radio treatment $(a=1)$ in comparison to the use of hormone treatment alone $(a=0)$. Thus, there were 588 patients in this subgroup ($38.3\%$ patients with hormone- and- radio treatment vs $61.7\%$ with hormone treatment alone).  Supplementary Table S.5 summarizes the effects of covariates on the marginal survival distribution for each outcome. Figure \ref{fig:causal_breast2} displays causal estimands under the AD or ND stratum. Similar to the analysis above, we dropped the variable ER status due to very small estimated coefficients (Supplementary Table S.5). 

\textbf{Sensitivity analysis.} To more comprehensively evaluate the robustness of our findings to unmeasured confounding, we conducted sensitivity analysis by following the strategy in Section \ref{section:sensitivity}, \ref{sec:estimation_frailty} and varying the pre-specified $\sigma$ values. We also compared our method to the Kaplan–Meier analysis based upon the samples matched by propensity scores (PS) using baseline variables as covariates in the PS model \citep{austin2014use}, a widely used method in clinical studies. Supplementary Figure S.7 summarizes the results. 

\textbf{Result interpretation.} (1) Results from the Copula regression model (Table S.4-S.5) indicated that there was a strongly positive association between RFS and OS even after adjusting for covariates. This highlights the importance of modeling dependence between semi-parametric risks. These tables also indicated that the impacts of covariates (such as Her2, Node, Size) on RFS ($T_1$) and OS ($T_2$) varied between two scenarios (i.e., $a=1$ vs $a=0$). However, sensitivity analyses, involving variations in the pre-specified $\sigma$ values, suggested that the frailty variance may primarily affect the association between $T_1$ and $T_2$ and has a lesser influence on the significance of covariates.  Specifically, in the first primary analysis, only the NPI variable consistently showed a positive association with $T_1$ for patients with $a=0$, while the variables Node, NPI, and Size were positively associated with $T_1$ for patients with $a=1$. In the second primary analysis, the variables Node, NPI, and Size were positively associated with $T_1$ for patients with $a=0$, while HER2, Size, and Meno had an impact for patients with $a=1$, where the first two showed a positive association and the last one showed a negative association. 

(2) Results from causal estimand estimation: by assuming no unmeasured confounding (i.e, $\sigma=0$), using Hormone therapy alone compared to no treatment has a positive impact and is beneficial significantly to those patients in ``ND" stratum at some duration after surgery. But it may not have significant effects on both RFS and OS for patients in ``AD" stratum with long-time treatment (Figure \ref{fig:causal_breast}). These imply that using Hormone therapy alone could be good for patients' overall survival in ``ND" stratum, compared to no treatment use. Sensitivity analysis with varied $\sigma$ also supported the above conclusion (Figure \ref{fig:causal_breast}). On the other hand, we detected a weak effect by comparing combined therapy with Hormone therapy alone (Figure \ref{fig:causal_breast2}) in ``AD" stratum at some duration after surgery. This implies that using the combined therapy might be even better for patients' overall survival in ``AD" stratum, compared to using Hormone therapy alone. 

(3) In the Kaplan–Meier analysis under PS-matched cohorts (Supplementary Figure S.7), we observed that not undergoing any treatment resulted in a significantly higher survival rate compared to hormone treatment, and the combined treatment did not exhibit a clear effect on OS, contrary to expectations in clinical practice. As for the KM survival curves of RFS, the combined treatment exhibited a significant effect on postponing the occurrence of relapse. This discrepancy could be attributed to the Kaplan–Meier analysis, even after PS matching, ignoring the dependence structure between RFS and OS and failing to account for potential differences over principal strata. Consequently, we place more emphasis on and advocate for the use of our proposed methods, which offer a more valid causal comparison and yield more reasonable final results.

\section{Simulation studies}
\label{sec:simulation}


The data generation mechanisms are classified into two categories: the ``without frailty" class (where the cross-world dependence is exclusively captured by all observed covariates) and the ``with frailty" class (where the cross-world dependence is not only attributed to observed covariates but also to latent and unobservable frailty variables). 
We evaluated the proposed methods based on four simulation settings (Ex1-Ex4 cases), where case Ex2 contains more covariates than case Ex1; case Ex3 is used to examine the robustness of our causal framework when treatment allocation is affected by baseline covariates as indicated by Assumption 2; and case Ex4 is to evaluate our method in the presence of a frailty describing two-world dependency. Due to the page limit, we kindly refer readers to Supplementary Section 4 for more technical details.

Under each scenario, we generated $200$ Monte Carlo replicates with sample size $n=500$ or $1000$. In all scenarios, we first estimated regression coefficients, baseline cumulative hazard functions, and the association between nonterminal and terminal times for each arm.  We provide a comprehensive evaluation of the estimation results of regression coefficients and association parameters in terms of mean biases, Monte Carlo standard deviations (MCSD), and asymptotic standard errors (ASE) across 200 simulation runs. We only present the results from case Ex1. The other three cases show similar patterns, thus omitted in this paper. We then estimated stratum-specific survivor causal effects \eqref{eq:SCE1}-\eqref{eq:SCE3} and evaluated their performance under cases Ex1-Ex4 varying with time over a grid of 30 time points which is determined by the sample order statistics of observed $T_1$. We used the bootstrap method with 100 replicates to estimate the standard errors of the estimators for both causal parameters and regression coefficients.

\begin{figure}[h]
\includegraphics[height=9cm,width=16cm]{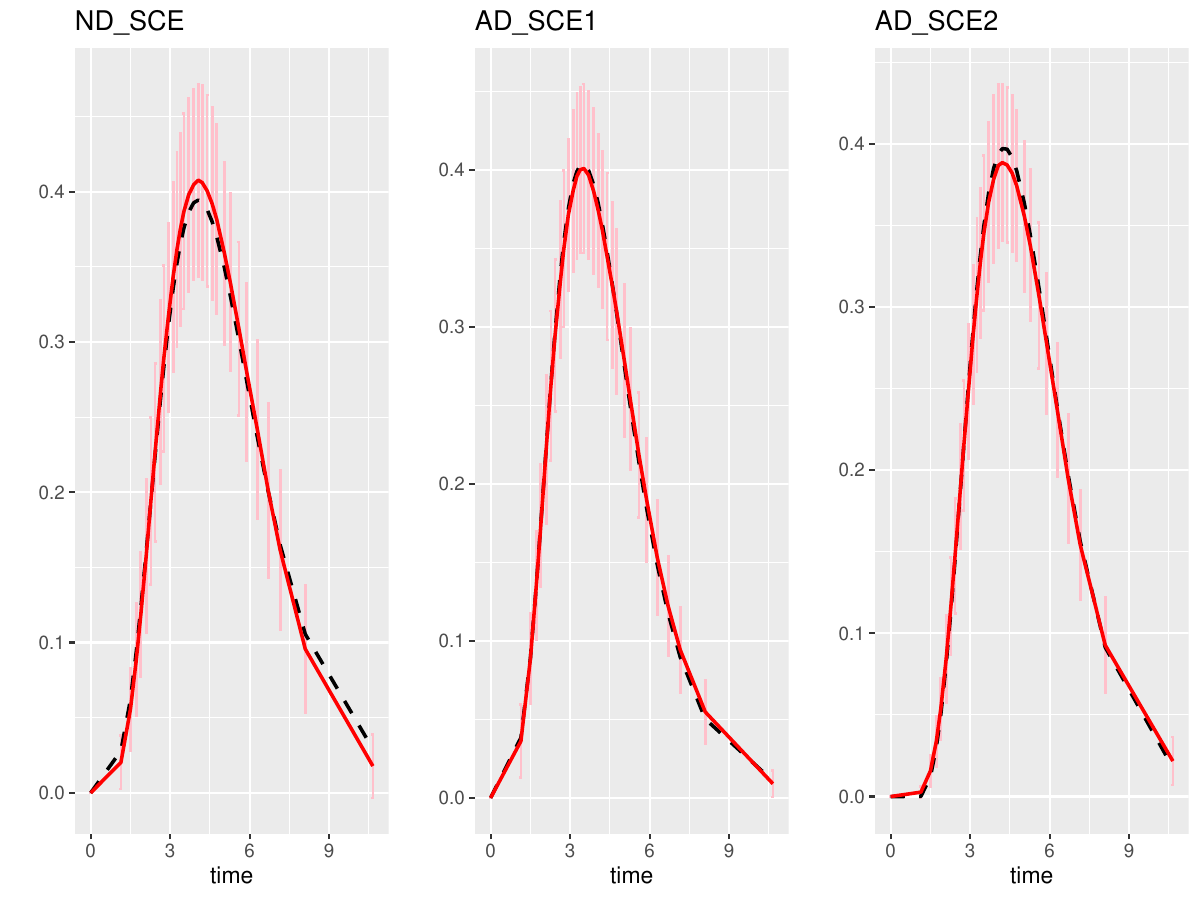}
\includegraphics[height=9cm,width=16cm]{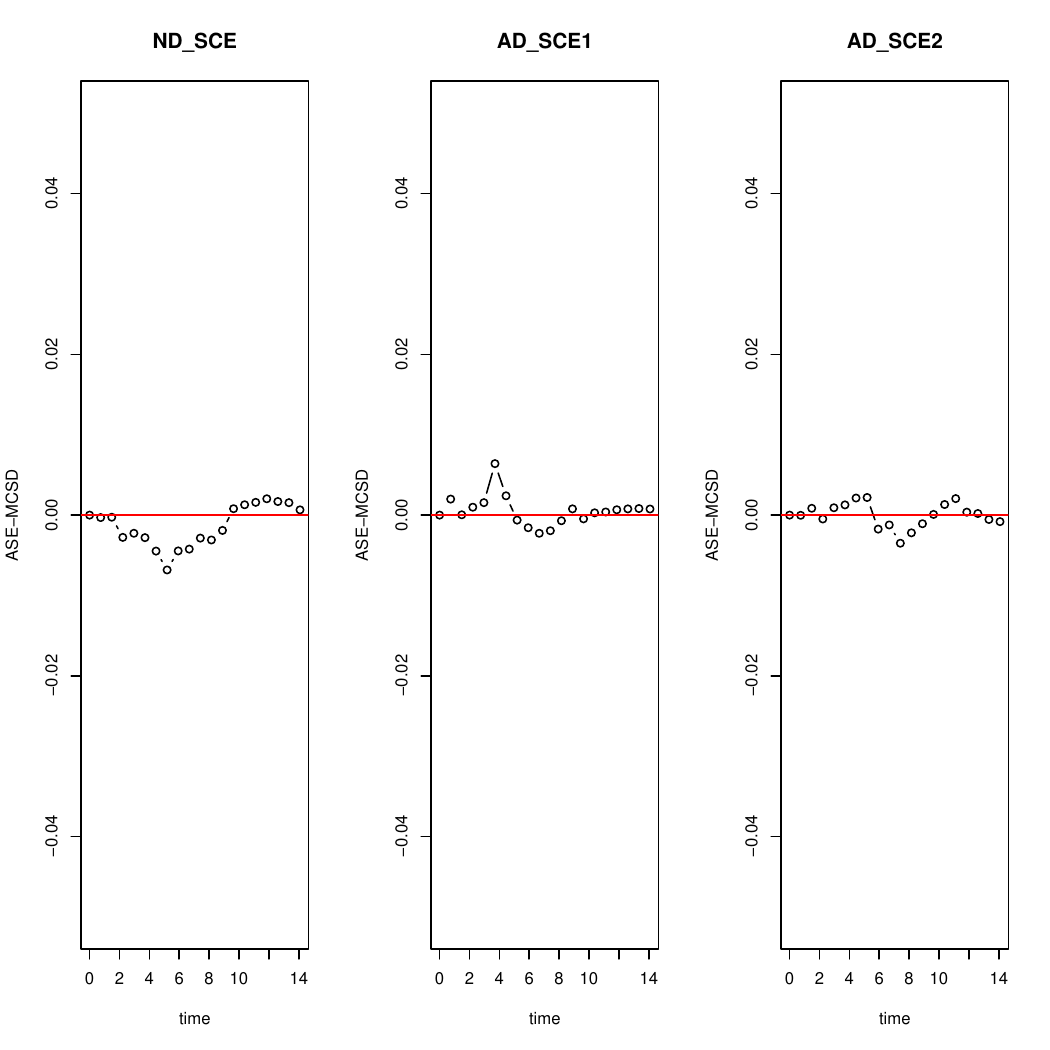}
\caption{Estimation of causal parameters under the case Ex1 with $\tau=0.3$ and $n=1000$. The three plots in the top row present the mean estimated causal parameters (represented by red curves) plus/minus 1.96 times the empirical standard deviation, encapsulating the true values (depicted in black dashed curves). The three plots in the bottom row present the difference between Boostrapping-based ASE and MCSD fluctuating around zero. From left to right, these plots correspond to $\operatorname{ND-SCE_2}$,$\operatorname{AD-SCE_1}$ and $\operatorname{AD-SCE_2}$, respectively.}
\label{fig:causal0.3_ex1}
\end{figure}

\begin{figure}[h]
\includegraphics[height=9cm,width=17cm]{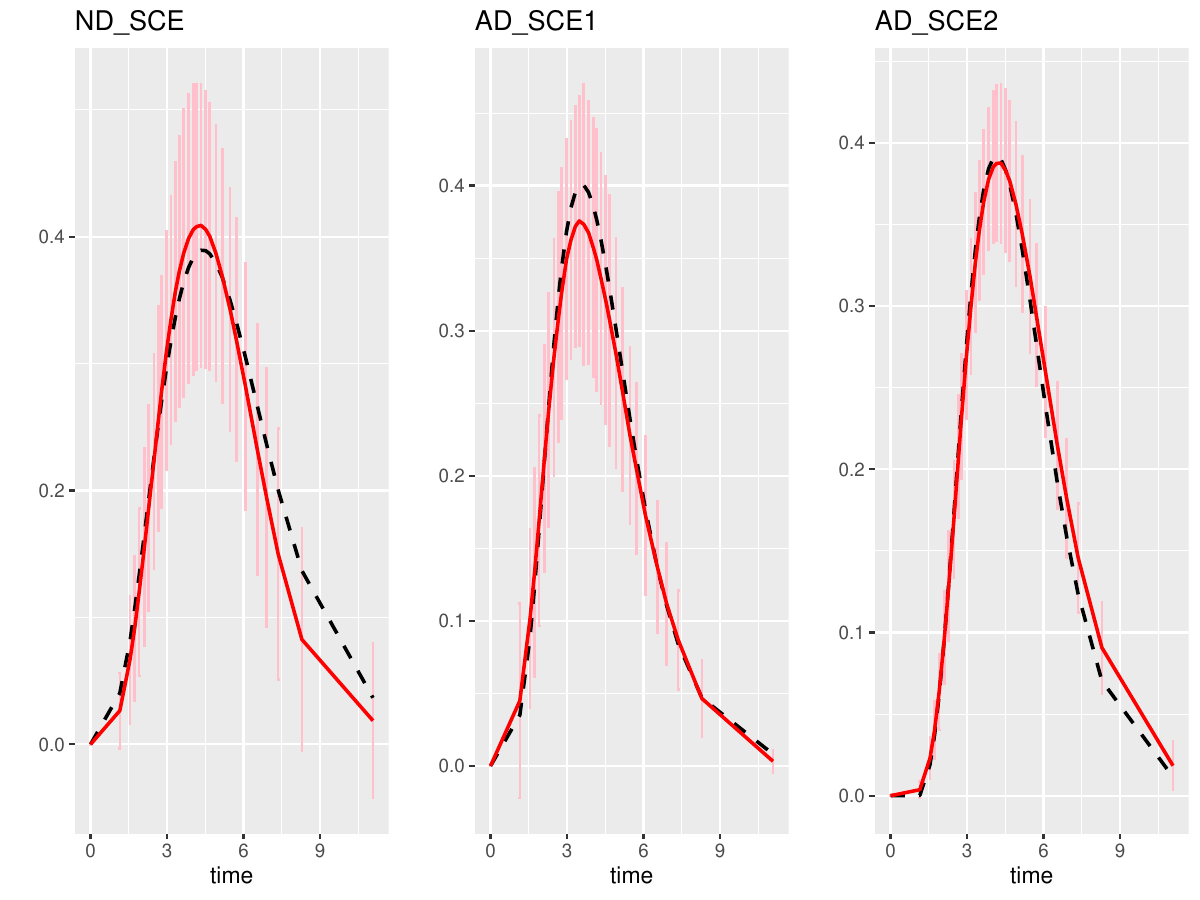}
\includegraphics[height=9cm,width=17cm]{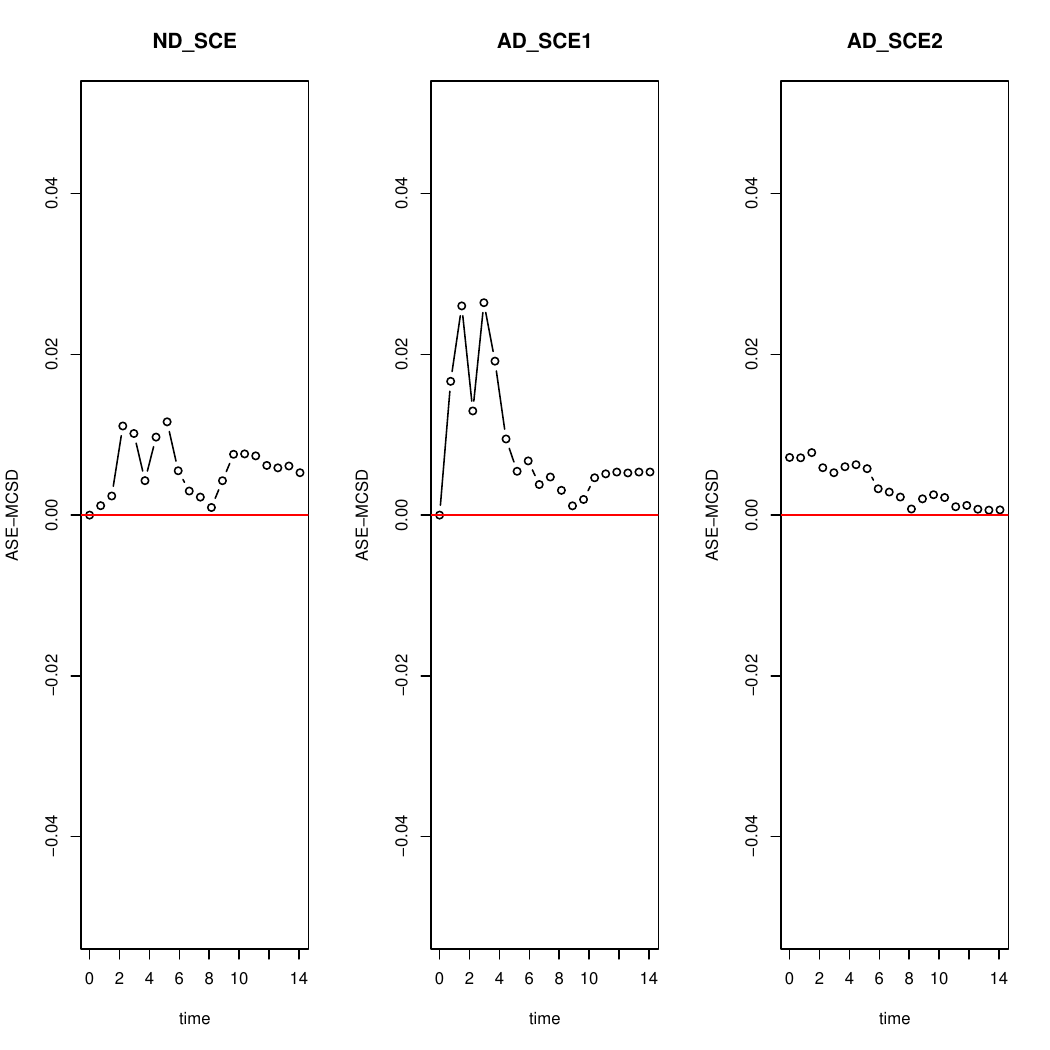}
\caption{Estimation of causal parameters under the case Ex1 with $\tau=0.6$ and $n=1000$. The three plots in the top row present the mean estimated causal parameters (represented by red curves) plus/minus 1.96 times the empirical standard deviation, encapsulating the true values (depicted in black dashed curves). The three plots in the bottom row present the difference between Boostrapping-based ASE and MCSD fluctuating around zero. From left to right, these plots correspond to $\operatorname{ND-SCE_2}$,$\operatorname{AD-SCE_1}$ and $\operatorname{AD-SCE_2}$, respectively.}
    \label{fig:causal0.6_ex1}
\end{figure}

Under Scenario Ex1 with low/high censoring setting, the biases and MCSDs of estimated regression coefficients appear to be negligible and diminish as the sample size increases for both Kendall's tau values (0.3 and 0.6) (Supplementary Table S.2). The estimated standard errors are close to the empirical standard deviation of point estimates of regression coefficients. The estimated baseline cumulative hazard functions (depicted in red) for the first 50 simulated samples surround the truth in blue (Supplementary Figure S.1).  In Figures \ref{fig:causal0.3_ex1}-\ref{fig:causal0.6_ex1}, the mean estimated causal parameters (red curves)  plus/minus 1.96 times empirical standard deviation (pink vertical bars) encapsulate the truth (depicted in black dashed curves), and discrepancies between ASE and MCSD at 20 time points seem to be negligible and fluctuate around zero. In Table \ref{tab:est_sce_cp}, biases of estimated causal parameters at some specific time points are small and their coverage probabilities (CP) are reasonable but slightly fluctuated around the nominal level 0.95, which is expected regarding the complex nature of causal estimand estimation.

\begin{figure}[h]
  \medskip

  \medskip
\includegraphics[height=9cm,width=17cm]{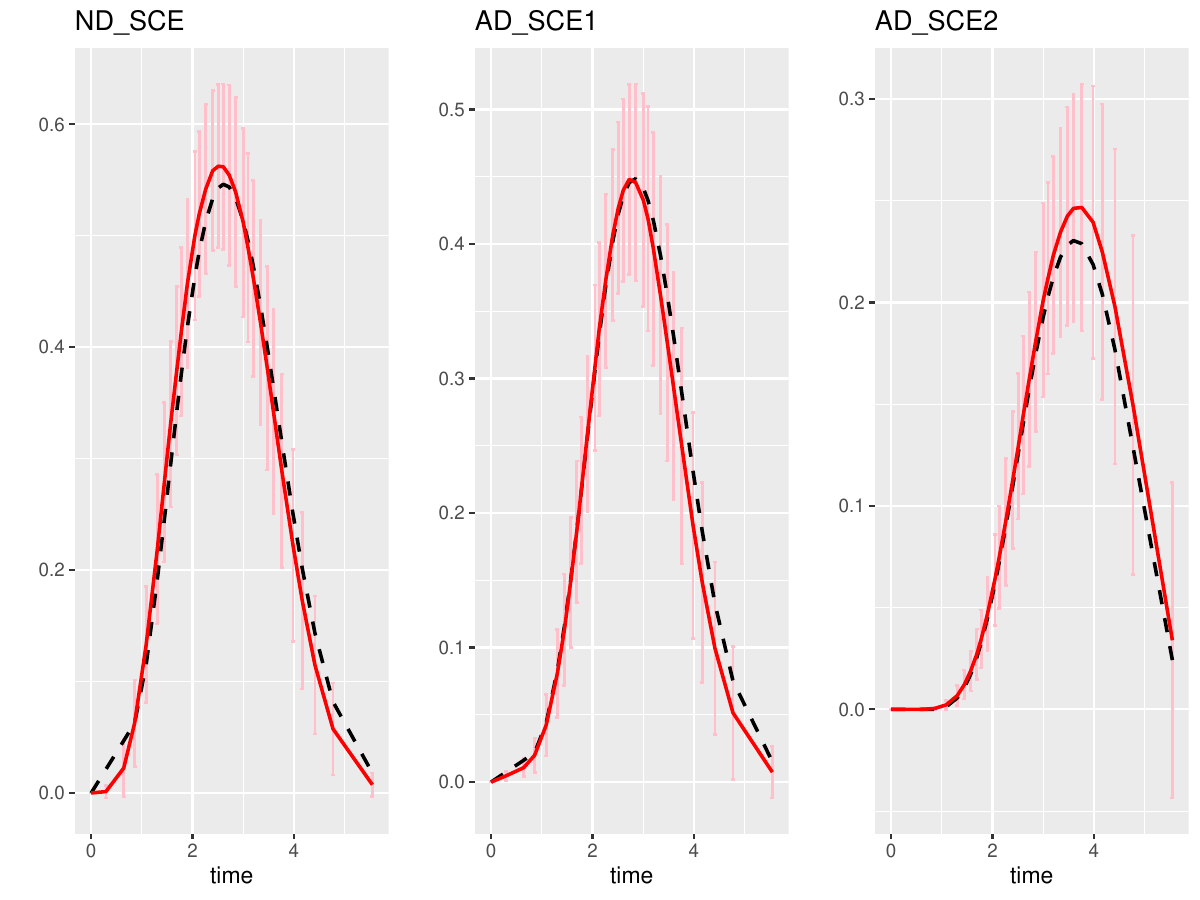}
\begin{center}
    \texttt{(a) $\sigma = 0.2$}
\end{center}
\includegraphics[height=9cm,width=17cm]{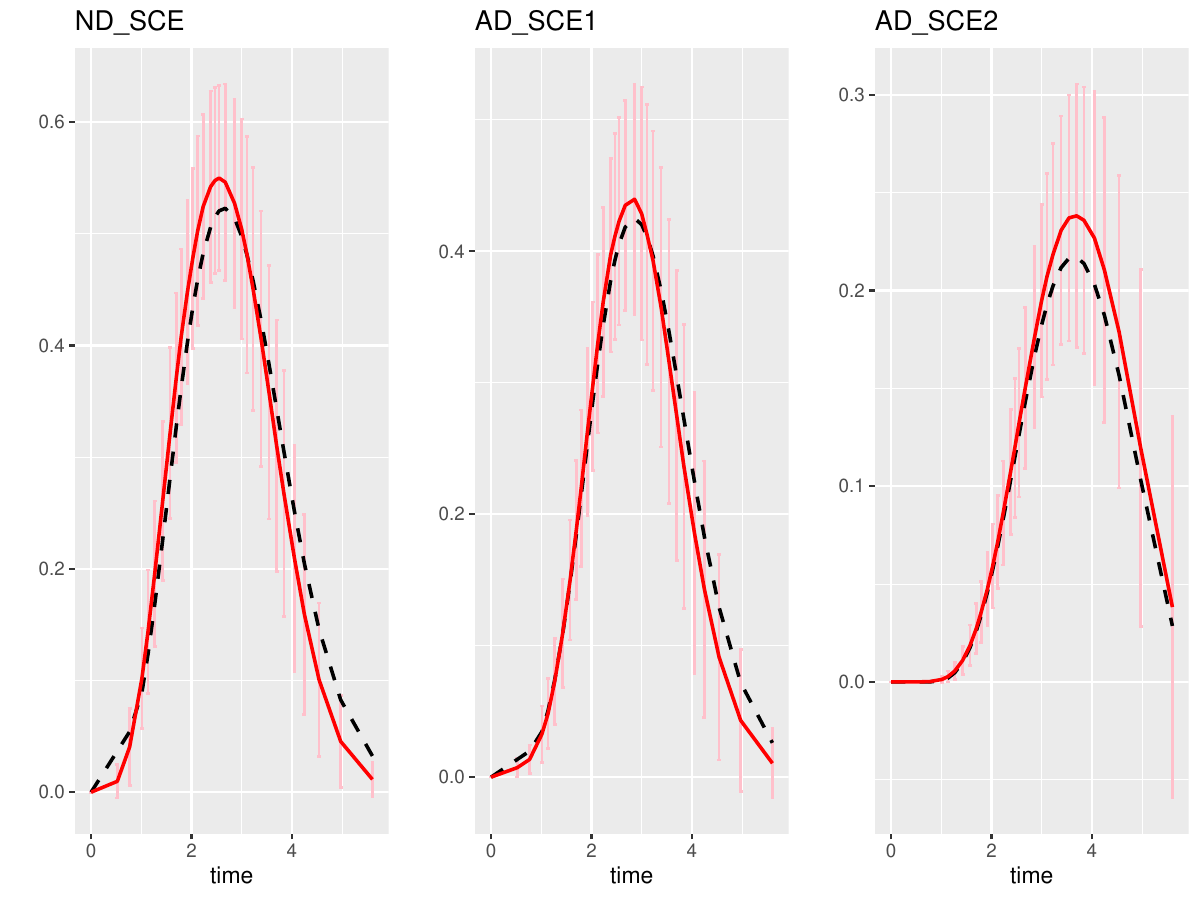}
\begin{center}
    \texttt{(b) $\sigma = 0.4$}
\end{center}

\caption{Estimation of causal parameters ($\operatorname{ND-SCE_2}$,$\operatorname{AD-SCE_1}$ and $\operatorname{AD-SCE_2}$) in the case Ex4 (evaluating the method for sensitivity analysis) with $n=1000$, $\tau=0.6$ and $\sigma=0.2/0.4$.}
    \label{fig:causal0.6_ex41}
\end{figure}

In the presence of more covariates as in case Ex2, estimators of causal parameters still perform well in terms of negligible estimation bias (Supplementary Figure S.2). Under case Ex3, it can be seen from Figure S.3 in the supplementary material that the proposed estimator shows satisfactory performance except for a slight deviation shown in the ND stratum. 
This anomaly is understandable in light of the high censoring environment, where fewer than $8\%$ of subjects belong to the ND stratum who are observed to have the $T_2$ event. 
In case Ex4 with a frailty describing two-world dependence, the proposed MCEM algorithm in Section \ref{section:estimation} yields asymptotically unbiased estimation for regression coefficients (Supplementary Table S.3) as well as causal parameters (Figure \ref{fig:causal0.6_ex41}). The results presented in the manuscript are based on $n=1000$, Kendall's $\tau=0.6$, and $\sigma = 0.2$ or $0.4$. More simulation evaluations based upon different combinations of sample size $n$, Kendall's $\tau$, and $\sigma$ can be found in Supplementary Figure S.4-S.6. Similar patterns are observed over all the settings, demonstrating the validity of our new method.

\section{Discussion}
\label{sec:conclusion}

This article establishes a frequentist framework using copula models for causal interpretation, net quantity estimation, and sensitivity analysis for unmeasured factors under right-censored semi-competing risks data. We have demonstrated the non-parametric identification and justified the feasibility and utility of the proposed estimation. Moreover, by applying our method, we detected that hormone therapy could be beneficial to patients with a low risk of relapse compared to no treatment use, and that the combined treatment will be beneficial to those with a higher risk of relapse, compared to hormone therapy alone. These findings imply a more tailored intervention strategy and suggest intermediate intervention after assessing the risk of relapse could be much more beneficial to patients' overall survival, which to some extent aligns with previous clinical research \citep{sledge2014} and provides more insight into personalized medicine for breast cancer patients to improve their post-surgery outcomes. 

For future method development, we introduce various extensions and pose open questions for readers to explore and enhance the current framework:  (1) Copula model selection. \cite{hsieh2008} provided the hypothesis test in terms of whether a copula model fits the data. Specifically, for each covariate group, the function $G_a(t_1,t_2) = \Pr(T_1\geq t_1, T_2\geq t_2|\delta_1 = 1,\delta_2 = 1, A=a)$ is identifiable non-parametrically in the upper wedge as
$\widetilde{G}_a (t_1,t_2)= \sum\limits_{i=1}^n I(X_i\geq t_1, Y_i\geq t_2,\delta_{i1}=1,\delta_{i2}=1,A_i=a)/\sum\limits_{i=1}^n I(\delta_{i1}=1,\delta_{i2}=1,A_i=a).$
The goodness-of-fit test can be thus constructed by comparing $\widetilde{G}_a(t_1,t_2)$ and $\widehat{G}_a(t_1,t_2)$ based on two copula models on the basis of some distance measure \citep{hsieh2008}. (2) This paper considers a simple copula model with homogeneous association for illustration.  The extensions of such a model include time-dependent association as \cite{peng2007semicompeting} and covariate-dependent association as in \cite{li2015}. (3) Another extension is to study multiple treatments
involving considering the effects of various interventions on survival outcomes subject to semi-competing risks while addressing potential confounding factors. This extension often requires methodologies like the generalized propensity score, instrumental variables, or causal mediation analysis adapted to handle multiple treatment groups. 
All of the above merits further investigation in future studies.




\bibliography{ref_causal}
\bibliographystyle{apalike}
 
\section*{Supporting Information}
The supplementary material contains technical proofs of propositions referenced in Section 3, the maximum likelihood estimation algorithm referenced in Section 4, and additional numerical results referenced in Sections 5-6.


\begin{table}[htbp]
\small
\caption{Estimation of causal parameters ($\operatorname{ND-SCE_2}$,$\operatorname{AD-SCE_1}$ and $\operatorname{AD-SCE_2}$) at time points $t=3,6$ in the case Ex1 with low censoring. The true values of causal parameters, the mean biases, and the coverage probabilities (CP) of their estimates are shown in this table.}
\label{tab:est_sce_cp}
\begin{center}
\begin{tabular}{rrrrrrrrrrrrrr}
\hline\hline
\multicolumn{1}{c}{}&\multicolumn{1}{c}{}&\multicolumn{3}{c}{$\operatorname{ND-SCE_2}$}&\multicolumn{1}{c}{}&\multicolumn{3}{c}{$\operatorname{AD-SCE_1}$ }&\multicolumn{1}{c}{}&\multicolumn{3}{c}{$\operatorname{AD-SCE_2}$}\tabularnewline
\cline{3-5}\cline{7-9}\cline{11-13}
Time& &Truth&Bias&95\%CP& &Truth&Bias&95\%CP& &Truth&Bias&95\%CP\tabularnewline
\hline
\multicolumn{5}{l}{$\tau=0.3,n=500$}\tabularnewline
$3$&&$0.32$&$-0.002$&$0.96$&&$0.38$&$0.003$&$1.00$&&$0.26$&$-0.004$&$0.94$\tabularnewline
$6$&&$0.26$&$ 0.013$&$0.96$&&$0.17$&$0.009$&$0.94$&&$0.27$&$-0.005$&$0.96$\tabularnewline
\hline
\multicolumn{5}{l}{$\tau=0.3, n=1000$}\tabularnewline
$3$&&$0.32$&$0.005$&$0.94$&&$0.38$&$-0.008$&$0.92$&&$0.26$&$-0.004$&$1.00$\tabularnewline
$6$&&$0.26$&$0.004$&$0.92$&&$0.17$&$ 0.002$&$0.90$&&$0.27$&$-0.006$&$0.94$\tabularnewline
\hline
\multicolumn{5}{l}{$\tau=0.6, n=500$}\tabularnewline
3&&$0.30$&$ 0.018$&$0.98$&&$0.37$&$-0.026$&$0.98$&&$0.28$&$-0.010$&$0.98$\tabularnewline
$6$&&$0.31$&$-0.055$&$0.94$&&$0.18$&$-0.017$&$1.00$&&$0.25$&$ 0.028$&$0.86$\tabularnewline
\hline
\multicolumn{5}{l}{$\tau=0.6, n=1000$}\tabularnewline
$3$&&$0.30$&$ 0.020$&$0.96$&&$0.37$&$-0.031$&$0.98$&&$0.28$&$-0.011$&$1.00$\tabularnewline
$6$&&$0.31$&$-0.015$&$0.94$&&$0.18$&$-0.010$&$0.98$&&$0.25$&$ 0.016$&$0.92$\tabularnewline
\hline
\end{tabular}\end{center}
\end{table}

\end{document}


\label{firstpage}

 \begin{center}
    {\LARGE\bf Supplementary Materials for ``Exploring causal effects of hormone- and radio-treatments in an observational study of breast cancer using copula-based semi-competing risks models"}
\end{center}
 \begin{center}
 \large Tonghui Yu$^1$, Mengjiao Peng$^2$, Yifan Cui$^3$ ,Elynn Chen$^{4}$, Chixiang Chen$^{5,6,*}$\\
 \end{center}
 \begin{center}
	$^{1}$
   School of Physical and Mathematical Sciences \\
   Nanyang Technological University, Singapore\\ 
	$^{2}$
    School of Statistics, East China Normal University, China\\ 
	$^{3}$
    Center for Data Science, Zhejiang University, China\\ 
    $^4$  Stern School of Business, New York University, New York, NY, USA.\\
    $^5$  Division of Biostatistics and Bioinformatics,\\ Department of Epidemiology and Public Health,\\ University of Maryland School of Medicine, Baltimore, MD, U.S.A.\\
    $^6$ Department of Neurosurgery,\\
    University of Maryland School of Medicine, Baltimore, MD, USA
    \end{center}
     \begin{center}
	{$^{*}$\textit{Contact Email: chixiang.chen@som.umaryland.edu}}
 \end{center}

Section \ref{sec:proof} provides proof of Proposition 1 from the main context. Section \ref{sec:prop2} gives the identification of causal parameters in the presence of unmeasured confounding. Section \ref{sec:mcem algorithm} summarizes the MCEM algorithm, used in Section 4.2 from the main context, for estimating regression parameters in the presence of unmeasured confounding. Section \ref{sec:extra simulation} contains more extra results in simulation studies. Section \ref{sec:extra real} 
provides additional results for the breast cancer study. 

\section{Proof of Proposition 1}
\label{sec:proof}
\textbf{Proof.}
Note that from the law of total probability
\begin{equation}
\Pr(T_k(a)\geq t|ad) 
= \int_0^{\infty} \Pr(T_k(a)\geq t|T_1(0)\leq T_2(0), T_1(1)\leq T_2(1),\bfZ) f_{\bfZ}(\bfZ|ad)d\bfZ.
\label{eq:pf_ad1}
\end{equation}
The integrand in \eqref{eq:pf_ad1} can be written as
\begin{equation*}
\begin{split}
&\Pr(T_k(a)\geq t|T_1(0)\leq T_2(0),T_1(1)\leq T_2(1),\bfZ) \\
& =  \Pr(T_k(a)\geq t|T_1(a)\leq T_2(a),\bfZ) \\
& = \Pr(T_k(a)\geq t|T_1(a)\leq T_2(a),A=a,\bfZ) \\
& = \Pr(T_k\geq t|T_1\leq T_2,A=a,\bfZ) :=S_{1|T_1\leq T_2,A=a,\bfZ}(t),
\end{split}
\end{equation*}
where the first equality is from Assumption 3, the third by randomization in Assumption 2, and the fourth by Assumption 1. Besides, the term $f_{\bfZ}(\bfZ|ad)$ in \eqref{eq:pf_ad1} can be written as
\begin{equation*}
\begin{split}
f_{\bfZ|ad}(\bfZ) 
& =  \frac{\Pr( T_1(0)\leq T_2(0), T_1(1)\leq T_2(1)|\bfZ)f_{\bfZ}(\bfZ)}{\int_0^{\infty} \Pr( T_1(0)\leq T_2(0), T_1(1)\leq T_2(1)|\bfZ)f_{\bfZ}(\bfZ)d\bfZ}\\
& =  \frac{\Pr( T_1(0)\leq T_2(0)|\bfZ)\Pr( T_1(1)\leq T_2(1)|\bfZ)f_{\bfZ}(\bfZ)}{\int_0^{\infty} \Pr( T_1(0)\leq T_2(0)|\bfZ)\Pr( T_1(1)\leq T_2(1)|\bfZ)f_{\bfZ}(\bfZ)d\bfZ}\\
& = \frac{\Pi_{A=0,\bfZ}\Pi_{A=1,\bfZ}f_{\bfZ}(\bfZ)}{\int_0^{\infty} \Pi_{A=0,\bfZ}\Pi_{A=1,\bfZ}f_{\bfZ}(\bfZ)d\bfZ}
\end{split}
\end{equation*}
where the first equality is by the Bayes' theorem, the second by Assumption 3, and the third by randomization in Assumptions 1-2. The proof for the identification of $\operatorname{ND-SCE}$ is similar.
Thus the results in Proposition 1 can be directly obtained.

\section{Identification of causal parameters with unmeasured confounding}
\label{sec:prop2}
\begin{prop}
\label{prop:scae}
Under Assumptions 1-2, 5-6, the stratum-specific survivor average causal effects in (4)-(6) in the main context 
are identified by
\begin{equation}
\begin{split}
&\operatorname{AD-SCE}_1(t)\\
& = \frac{\mathbb{E}_{\gamma,\bfZ}\left\{ \int_t^{\infty} D_1 \left(s,s|1,\gamma,\bfZ  \right)d S_1(s| 1,\gamma,\bfZ ) \int_0^{\infty} D_1 \left(s,s|0,\gamma,\bfZ  \right)d S_1(s| 0,\gamma,\bfZ )
\right\}}{\mathbb{E}_{\gamma,\bfZ}\left\{ 
\int_0^{\infty} D_1 \left(s,s|1,\gamma,\bfZ  \right)d S_1(s| 1,\gamma,\bfZ ) \int_0^{\infty} D_1 \left(s,s|0,\gamma,\bfZ  \right)d S_1(s| 0,\gamma,\bfZ )\right\}}\\
&- \frac{\mathbb{E}_{\gamma,\bfZ}\left\{ 
\int_t^{\infty} D_1 \left(s,s|0,\gamma,\bfZ  \right)d S_1(s| 0,\gamma,\bfZ )
  \int_0^{\infty} D_1 \left(s,s|1,\gamma,\bfZ  \right)d S_1(s| 1,\gamma,\bfZ )
  \right\}}{\mathbb{E}_{\gamma,\bfZ}\left\{ 
  \int_0^{\infty} D_1 \left(s,s|1,\gamma,\bfZ  \right)d S_1(s| 1,\gamma,\bfZ )\int_0^{\infty} D_1 \left(s,s|0,\gamma,\bfZ  \right)d S_1(s| 0,\gamma,\bfZ )\right\}},
\end{split}
\label{eq:adSCE_f1}
\end{equation}
\begin{equation}
\begin{split}
&\operatorname{AD-SCE}_2(t)\\
& = \frac{\mathbb{E}_{\gamma,\bfZ}\left\{ \left[
S_2(t| 1,\gamma,\bfZ )+\int_t^{\infty}D_2 \left(s,s|1,\gamma,\bfZ  \right)d S_2(s| 1,\gamma,\bfZ )\right]\left[1+\int_0^{\infty}D_2 \left(s,s|0,\gamma,\bfZ  \right)d S_2(s| 0,\gamma,\bfZ )\right]
  \right\}}{\mathbb{E}_{\gamma,\bfZ}\left\{ 
\left[1+\int_0^{\infty}D_2 \left(s,s|0,\gamma,\bfZ  \right)d S_2(s| 0,\gamma,\bfZ )\right]\left[1+\int_0^{\infty}D_2 \left(s,s|1,\gamma,\bfZ  \right)d S_2(s| 1,\gamma,\bfZ )\right] \right\}}\\
&- \frac{\mathbb{E}_{\gamma,\bfZ}\left\{ \left[
S_2(t| 0,\gamma,\bfZ )+\int_t^{\infty}D_2 \left(s,s|0,\gamma,\bfZ  \right)d S_2(s| 0,\gamma,\bfZ )\right]\left[1+\int_0^{\infty}D_2 \left(s,s|1,\gamma,\bfZ  \right)d S_2(s| 1,\gamma,\bfZ )\right]
  \right\}}{\mathbb{E}_{\gamma,\bfZ}\left\{ 
\left[1+\int_0^{\infty}D_2 \left(s,s|0,\gamma,\bfZ  \right)d S_2(s| 0,\gamma,\bfZ )\right]\left[1+\int_0^{\infty}D_2 \left(s,s|1,\gamma,\bfZ  \right)d S_2(s| 1,\gamma,\bfZ )\right] \right\}},
\end{split}
\label{eq:adSCE_f2}
\end{equation}
\begin{equation}
\begin{split}
&\operatorname{ND-SCE}_2(t)\\
& = \frac{\mathbb{E}_{\gamma,\bfZ}\left\{ \int_t^{\infty} D_2 \left(s,s|1,\gamma,\bfZ  \right)d S_2(s| 1,\gamma,\bfZ ) \int_0^{\infty} D_2 \left(s,s|0,\gamma,\bfZ  \right)d S_2(s| 0,\gamma,\bfZ )
\right\}}{\mathbb{E}_{\gamma,\bfZ}\left\{ 
\int_0^{\infty} D_2 \left(s,s|1,\gamma,\bfZ  \right)d S_2(s| 1,\gamma,\bfZ ) \int_0^{\infty} D_2 \left(s,s|0,\gamma,\bfZ  \right)d S_2(s| 0,\gamma,\bfZ )\right\}}\\
&- \frac{\mathbb{E}_{\gamma,\bfZ}\left\{ 
\int_t^{\infty} D_2 \left(s,s|0,\gamma,\bfZ  \right)d S_2(s| 0,\gamma,\bfZ )
  \int_0^{\infty} D_2 \left(s,s|1,\gamma,\bfZ  \right)d S_2(s| 1,\gamma,\bfZ )
  \right\}}{\mathbb{E}_{\gamma,\bfZ}\left\{ 
  \int_0^{\infty} D_2 \left(s,s|1,\gamma,\bfZ  \right)d S_2(s| 1,\gamma,\bfZ )\int_0^{\infty} D_2 \left(s,s|0,\gamma,\bfZ  \right)d S_2(s| 0,\gamma,\bfZ )\right\}}.
\end{split}
\label{eq:ndSCE_f}
\end{equation}
\end{prop}

\section{MCEM algorithm}
\label{sec:mcem algorithm}
The EM algorithm is a popular tool for maximizing likelihood functions in the presence of unobserved data. The E step is approximated using simulated samples from the exact conditional distribution of the unmeasured variables given the observed data, so the E-step is divided into a simulation and a MC integration step. Specifically, we first replace $\lambda_{0k}^{(a)}(t)$ 
with the jump sizes of $\Lambda_{0k}^{(a)}$ at time t denoted by $\Lambda_{0k}^{(a)}\{t\}$. The NPMLE estimates for $\Lambda_{01}^{(a)}\{t_1\}$ and $\Lambda_{02}^{(a)}\{t_2\}$ have the forms of
\begin{equation}
\widehat{\Lambda}_{01}^{(a)}\{t_1\} = \frac{\sum\limits_{j=1}^n \delta_{j1}I(A_j=a)I(X_j=t_1)}{\sum\limits_{j=1}^n I(X_j\geq t_1)I(A_j=a)\exp(\bfbeta_1^{(a)T}\bfZ_j)E_{\gamma}[\gamma_jw_{1j}(a)|\mathcal{O}_j]},
\end{equation}
\begin{equation}
\widehat{\Lambda}_{02}^{(a)}\{t_2\} = \frac{\sum\limits_{j=1}^n \delta_{j2}I(A_j=a)I(Y_j=t_2)}{\sum\limits_{j=1}^n I(Y_j\geq t_2)I(A_j=a)\exp(\bfbeta_2^{(a)T}\bfZ_j)E_{\gamma}[\gamma_jw_{2j}(a)|\mathcal{O}_j]},
\end{equation}
where 
\begin{equation*}
\begin{split}
w_{1i}(a) &= \delta_{i1}\delta_{i2}\left[ \frac{D_{121}(X_i,Y_i;\alpha^{(a)})S_1(X_i|\bfZ_i,a,\gamma_i)}{D_{12}(X_i,Y_i;\alpha^{(a)})}+1\right]\\
&+\delta_{i1}(1-\delta_{i2})\left[\frac{D_{11}(X_i,Y_i;\alpha^{(a)})S_1(X_i|\bfZ_i,a,\gamma_i)}{D_{1}(X_i,Y_i;\alpha^{(a)})}+1 \right]\\
&+(1-\delta_{i1})\delta_{i2}\frac{D_{12}(X_i,Y_i;\alpha^{(a)})S_1(X_i|\bfZ_i,a,\gamma_i)}{D_{2}(X_i,Y_i;\alpha^{(a)})}\\
&+(1-\delta_{i1})(1-\delta_{i2})\frac{D_{1}(X_i,Y_i;\alpha^{(a)})S_1(X_i|\bfZ_i,a,\gamma_i)}{D(X_i,Y_i;\alpha^{(a)})},
\end{split}
\end{equation*}

\begin{equation*}
\begin{split}
w_{2i}(a) &= \delta_{i1}\delta_{i2}\left[ \frac{D_{122}(X_i,Y_i;\alpha^{(a)})S_2(Y_i|\bfZ_i,a,\gamma_i)}{D_{12}(X_i,Y_i;\alpha^{(a)})}+1\right]\\
&+\delta_{i1}(1-\delta_{i2})\left[\frac{D_{12}(X_i,Y_i;\alpha^{(a)})S_2(Y_i|\bfZ_i,a,\gamma_i)}{D_{1}(X_i,Y_i;\alpha^{(a)})}+1 \right]\\
&+(1-\delta_{i1})\delta_{i2}\frac{D_{22}(X_i,Y_i;\alpha^{(a)})S_2(Y_i|\bfZ_i,a,\gamma_i)}{D_{2}(X_i,Y_i;\alpha^{(a)})}\\
&+(1-\delta_{i1})(1-\delta_{i2})\frac{D_{2}(X_i,Y_i;\alpha^{(a)})S_2(Y_i|\bfZ_i,a,\gamma_i)}{D(X_i,Y_i;\alpha^{(a)})}.
\end{split}
\end{equation*}
Thus the baseline cumulative hazard functions for the nonterminal and terminal events are estimated by the following piece-wise constant functions
\begin{equation}
\widehat{\Lambda}_{01}^{(a)}(t_1)= \sum\limits_{i: X_i\leq t_1} \widehat{\Lambda}_{01}^{(a)}\{X_i\},
\end{equation}
\begin{equation}
\widehat{\Lambda}_{02}^{(a)}(t_2)= \sum\limits_{i: Y_i\leq t_2} \widehat{\Lambda}_{02}^{(a)}\{Y_i\}.
\end{equation}

Plugging $\lambda_{0k}^{(a)}(t)$ and $\Lambda_{0k}^{(a)}(t)$ by $\widehat{\Lambda}_{0k}^{(a)}\{t\}$ and $\widehat{\Lambda}_{0k}^{(a)}(t)$, respectively, into the log-likelihood (15) in the main context 
yields the profile log-likelihood of $\bfvarrho$. Due to the existence of unmeasured factor and intractable integral, a Monte Carlo Expectation-Maximization (MCEM) approach can be used to estimate $\bfvarrho$. Specifically, the complete log-likelihood function of $(\alpha^{(a)},\Lambda_{01}^{(a)},\Lambda_{02}^{(a)},\bfbeta_{1}^{(a)},\bfbeta_{2}^{(a)},\sigma)$ based on $(\mathcal{O},\bfgamma)$ is given by 
\begin{equation}
\begin{split}
&l_c(\alpha^{(a)},\Lambda_{01}^{(a)},\Lambda_{02}^{(a)},\bfbeta_{1}^{(a)},\bfbeta_{2}^{(a)},\sigma|\mathcal{O},\bfgamma) \\
&= \sum\limits_{i=1}^n l_{ci}(\alpha^{(a)},\Lambda_{01}^{(a)},\Lambda_{02}^{(a)},\bfbeta_{1}^{(a)},\bfbeta_{2}^{(a)},\sigma|\mathcal{O}_i,\gamma_i) \\
&= \sum\limits_{i=1}^n l_i(\alpha^{(a)},\Lambda_{01}^{(a)},\Lambda_{02}^{(a)},\bfbeta_{1}^{(a)},\bfbeta_{2}^{(a)}|\mathcal{O}_i,\gamma_i)+f_{\gamma}(\gamma_i|\sigma)
\end{split}
\end{equation}
In the E-step, we compute the expectation of the complete log-likelihood based on observed data $\mathcal{O}$, pre-specified $\sigma$ and current updated estimates $(\widehat{\bfvarrho}^{(r)},\widehat{\bfLambda}^{(r)})$ as follows.
\begin{equation}
\begin{split}
E\left[l_c(\bfvarrho,\widehat{\bfLambda}^{(r)},\sigma|\mathcal{O},\bfgamma)|\mathcal{O},\widehat{\bfvarrho}^{(r)}\right]& = \sum\limits_{i=1}^n \int l_{ci}(\bfvarrho,\widehat{\bfLambda}^{(r)},\sigma|\mathcal{O}_i,\gamma_i) f(\gamma_i|\mathcal{O}_i;\widehat{\bfvarrho}^{(r)},\widehat{\bfLambda}^{(r)},\sigma) d\gamma_i\\
&\approx \frac{1}{m}\sum\limits_{t=1}^m l_c(\bfvarrho,\widehat{\bfLambda}^{(r)},\sigma|\mathcal{O},\bfgamma_t^{(r)})
\l:= Q(\bfvarrho|\widehat{\bfvarrho}^{(r)})
\end{split}
\end{equation}
for a large enough integer $m$, where $\bfgamma_t^{(r)}=(\gamma_{t1}^{(r)},\cdots, \gamma_{tn}^{(r)})$ can be obtained from a Markov chain Monte Carlo (MCMC) routine such as the Gibbs sampler or
Metropolis–Hastings algorithm with posterior distribution 
$f(\gamma_i|\mathcal{O}_i;\widehat{\bfvarrho}^{(r)},\widehat{\bfLambda}^{(r)},\sigma)$ in the form of
\begin{equation}
f(\gamma_i|\mathcal{O}_i;\widehat{\bfvarrho},\widehat{\bfLambda},\sigma) 
= \frac{\exp\left[l_i(\widehat{\bfvarrho},\widehat{\bfLambda}|\mathcal{O}_i,\gamma_i) \right]f_{\gamma}(\gamma_i|\sigma)}{\int \exp\left[l_i(\widehat{\bfvarrho},\widehat{\bfLambda}|\mathcal{O}_i,\gamma_i) \right]f_{\gamma}(\gamma_i|\sigma)d\gamma_i}.
\end{equation}
In the M-step, the estimates of $\bfvarrho$ is updated via $\widehat{\bfvarrho}^{(r+1)} = \max_{\bfvarrho}Q(\bfvarrho|\widehat{\bfvarrho}^{(r)}).  $
Although the Monte Carlo approximation provides a solution to overcome the intractable E-step, it also has a persistent MC error depending on the sample size. This can sometimes be overcome by starting with small value of $m$ and increasing $m$ with iteration step $r$. The rule for increasing MCMC sample sizes can be found in \cite{caffo2005ascent} to make the algorithm move closer to convergence.

\section{Simulation Setups and extra simulation results}
\label{sec:extra simulation}

\textbf{Class without frailty.}
In the first scenario (Ex1), there are two covariates:  $Z_{i1}\sim$ uniform$[-1,1]$ and $Z_{i2}\sim N(0,1)$, $i=1,\cdots,n$. The treatment variable, denoted by $A_i$, independently follows a Bernoulli distribution with a probability of 0.5. Regardless of the assigned treatment, we employ the same Frank copula function with Kendall's tau $\tau^{(0)}=\tau^{(1)}=\tau$ set at either 0.3 or 0.6, where the definition of Kendall's tau is introduced in (3) from the main context. 
The marginal distributions for the non-terminal and terminal event times follow Cox proportional hazards models with Weilbull distributions at baseline. Different values are assigned to the regression coefficients and baseline hazard function under distinct treatment allocations. Specific details are provided in Table S.1. 
The censoring variable is drawn from a Uniform$[0, c_u]$, where $c_u$ is chosen to achieve the desired censoring rates. In the case of low censoring, we set $c_u=45$, resulting in approximately 35$\%$ for $T_1$ and 10$\%$ for $T_2$ censored in the samples; while in the case of high censoring, $c_u=16$ is chosen, leading to censoring rates of approximately 45$\%$ for $T_1$ and 30$\%$ for $T_2$. 
In the second example (Ex2), we consider more set of covariates to capture cross-world dependence. The covariates, denoted as $Z_{ip}$ ($p=1,\cdots,6$), are independently generated from a standard normal distribution. The regression coefficients for these covariates are specified in Table S.1. 
Other aspects of the setup are in accord with Ex1 with $c_u=16$. In the third example (Ex3), the setup is the same as in Ex1 with $c_u=16$, except for the treatment variable $A_i$, which is generated from a logistic regression in which the success probability is given by $\{1+\exp(-Z_{i1})\}^{-1}$. 

\textbf{Class with frailty.} 
We continue to consider an example (Ex4) with two covariates and a frailty variable. The generation of covariates and the treatment variable are the same with the first example Ex1. The frailty variable follows a Gamma distribution with shape and rate equal to $1/\sigma$, where $\sigma$ is set at either 0.2 or 0.4. We plug the true value of $\sigma$ into estimation for illustrating the proposed method. In practice, these frailty models will be used in the sensitivity analysis with several pre-specified $\sigma$ values. The marginal distributions for the nonterminal and terminal event times follow the shared-frailty proportional hazards models (14) in the main context. 
The true values of regression coefficients are specified in Table S.1 within the supplementary material. The joint distribution between nonterminal and terminal event times follows the Frank copula structure with Kendall's tau $\tau^{(0)}=\tau^{(1)}=\tau$, set at either 0.3 or 0.6. The censoring variable is simulated from a Uniform$[0, 12]$, leading to the censoring rates of approximately 50$\%$ for $T_1$ and 30$\%$ for $T_2$. 

\begin{table}[h]
    \centering
        \caption{Parameter specification in the simulated samples }
    \label{tab:sim_param}
    \begin{tabular}{cccccccc}
    \hline \hline
    \multicolumn{1}{l}{Examples}&\multicolumn{1}{c}{}&\multicolumn{2}{c}{$T_1$}&&\multicolumn{2}{c}{$T_2$}\\
    \cline{3-4}\cline{6-7}
    &&$\Lambda_{01}^{(a)}(t)$& $\beta_{1}^{(a)}$&&$\Lambda_{02}^{(a)}(t)$&$\beta_{2}^{(a)}$\\
     \hline
    Ex1/Ex3& $a=0$     &  $(t/3.5)^5$& (1,2) && $(t/4)^{5.5}$ &(1,2) \\
    &$a=1$     &  $(t/5.5)^6$ & (0,2) && $(t/5.8)^{6.5}$ &(0.5,2) \\
    \hline
    Ex2& $a=0$     &  $(t/3)^4$& (2/3,2/3,2/3,0,0,0) && $(t/3.4)^{4.2}$ &(0,0,0,2,2,2) \\
    &$a=1$     &  $(t/4)^5$ & (1,1,1,0,0,0) && $(t/4.5)^{5.2}$ &(0,0,0,1,1,1) \\  
    \hline
    Ex4& $a=0$     &  $(t/3)^4$& (2,0) && $(t/3.4)^{4.2}$ &(0,2) \\
    &$a=1$     &  $(t/4)^5$ & (1,0) && $(t/4.5)^{5.2}$ &(0,1) \\                 
    \hline
    \end{tabular}
\end{table}

\begin{table}[htbp]
\footnotesize
\begin{center}
\caption{Estimation results of regression coefficients and association parameters in the first example Ex1. The mean biases, Monte Carlo standard derivation (MCSD) and asymptotic standard error (SE) of estimated parameters are summarized in this table.}
\label{tab:est_coef_ex1}
\begin{tabular}{lrrrrrrrrrrrrrrrr}
\hline\hline
& &\multicolumn{7}{c}{$\tau=0.3$}&&\multicolumn{7}{c}{$\tau=0.6$}\tabularnewline
\cline{3-9}\cline{11-17}
& &\multicolumn{3}{c}{$n=500$}&&\multicolumn{3}{c}{$n=1000$}&&\multicolumn{3}{c}{$n=500$}&&\multicolumn{3}{c}{$n=1000$}\tabularnewline
\cline{3-5}\cline{7-9}\cline{11-13}
\cline{15-17}
 && Bias& MCSD&SE& & Bias& MCSD&SE& & Bias& MCSD&SE&& Bias& MCSD&SE \tabularnewline
 \hline
\multicolumn{5}{l}{low censoring}\tabularnewline
$\tau^{(0)}$&& 0.016&0.048&0.044& & 0.007&0.033&0.032& &-0.074&0.050&0.082& &-0.030&0.053&0.057\tabularnewline
$\tau^{(1)}$&& 0.019&0.048&0.046& & 0.015&0.034&0.035&&-0.157&0.081&0.104& &-0.089&0.074&0.071\tabularnewline
$\beta_{11}^{(0)}$&&-0.012&0.146&0.152& & 0.005&0.103&0.102&& 0.015&0.135&0.142&& 0.028&0.096& 0.096\tabularnewline
$\beta_{12}^{(0)}$&&-0.028&0.150&0.157& &-0.011&0.097&0.108&& 0.027&0.142&0.149&& 0.030&0.096&0.107\tabularnewline
$\beta_{11}^{(1)}$&&-0.019&0.154&0.152& &-0.015&0.087&0.104&&-0.070&0.138&0.148&&-0.040&0.082&0.097\tabularnewline
$\beta_{12}^{(1)}$&&-0.016&0.169&0.175& &-0.032&0.117&0.117&& 0.027&0.155&0.168&& 0.015&0.110&0.114\tabularnewline
$\beta_{21}^{(0)}$&& 0.019&0.126&0.131& & 0.007&0.092&0.091&& 0.046&0.137&0.148&& 0.035&0.092&0.096\tabularnewline
$\beta_{22}^{(0)}$&& 0.011&0.128&0.132& & 0.010&0.087&0.091&& 0.078&0.134&0.147&& 0.059&0.092&0.101\tabularnewline
$\beta_{21}^{(1)}$&&-0.007&0.120&0.130& & 0.000&0.091&0.085&&-0.027&0.130&0.149&&-0.018&0.091&0.096\tabularnewline
$\beta_{22}^{(1)}$&& 0.010&0.137&0.139& & 0.004&0.083&0.089&& 0.074&0.128&0.150&& 0.053&0.088&0.100\tabularnewline
\hline
\multicolumn{5}{l}{high censoring}\tabularnewline
$\tau^{(0)}$&& 0.016&0.052&0.047 && 0.006&0.036&0.035 &&-0.081&0.067&0.085 &&-0.032&0.049&0.054\tabularnewline
$\tau^{(1)}$&& 0.027&0.052&0.051 && 0.019&0.039&0.040 &&-0.166&0.094&0.106 &&-0.090&0.069&0.071\tabularnewline
$\beta_{11}^{(0)}$&&-0.007&0.166&0.166 && 0.009&0.111&0.110 && 0.017&0.154& 0.156&& 0.037&0.115&0.102\tabularnewline
$\beta_{12}^{(0)}$&&-0.020&0.158&0.168 &&-0.006&0.105&0.115 && 0.031&0.149&0.161 && 0.039&0.106&0.110\tabularnewline
$\beta_{11}^{(1)}$&&-0.015&0.177&0.181 &&-0.011&0.107& 0.121&&-0.072&0.162&0.175 &&-0.039&0.102&0.115\tabularnewline
$\beta_{12}^{(1)}$&&-0.001&0.183&0.193 &&-0.027&0.125&0.130 && 0.050&0.172& 0.191&& 0.027&0.123&0.128\tabularnewline
$\beta_{21}^{(0)}$&& 0.022&0.141&0.145 && 0.009&0.103&0.099 && 0.057&0.161&0.165 && 0.041&0.104&0.106\tabularnewline
$\beta_{22}^{(0)}$&& 0.013&0.134&0.142 && 0.011&0.096&0.099 && 0.099&0.149&0.161 && 0.068&0.097&0.110\tabularnewline
$\beta_{21}^{(1)}$&& 0.003&0.134&0.154 &&-0.001&0.104&0.101 &&-0.011&0.150&0.177 &&-0.005&0.111&0.117\tabularnewline
$\beta_{22}^{(1)}$&& 0.018&0.163&0.162 && 0.004&0.097&0.104 && 0.102&0.156&0.173 && 0.071&0.104&0.118\tabularnewline
\hline
\end{tabular}\end{center}
\end{table}

\begin{figure}[h]
  \medskip
\includegraphics[height=9cm,width=17cm]{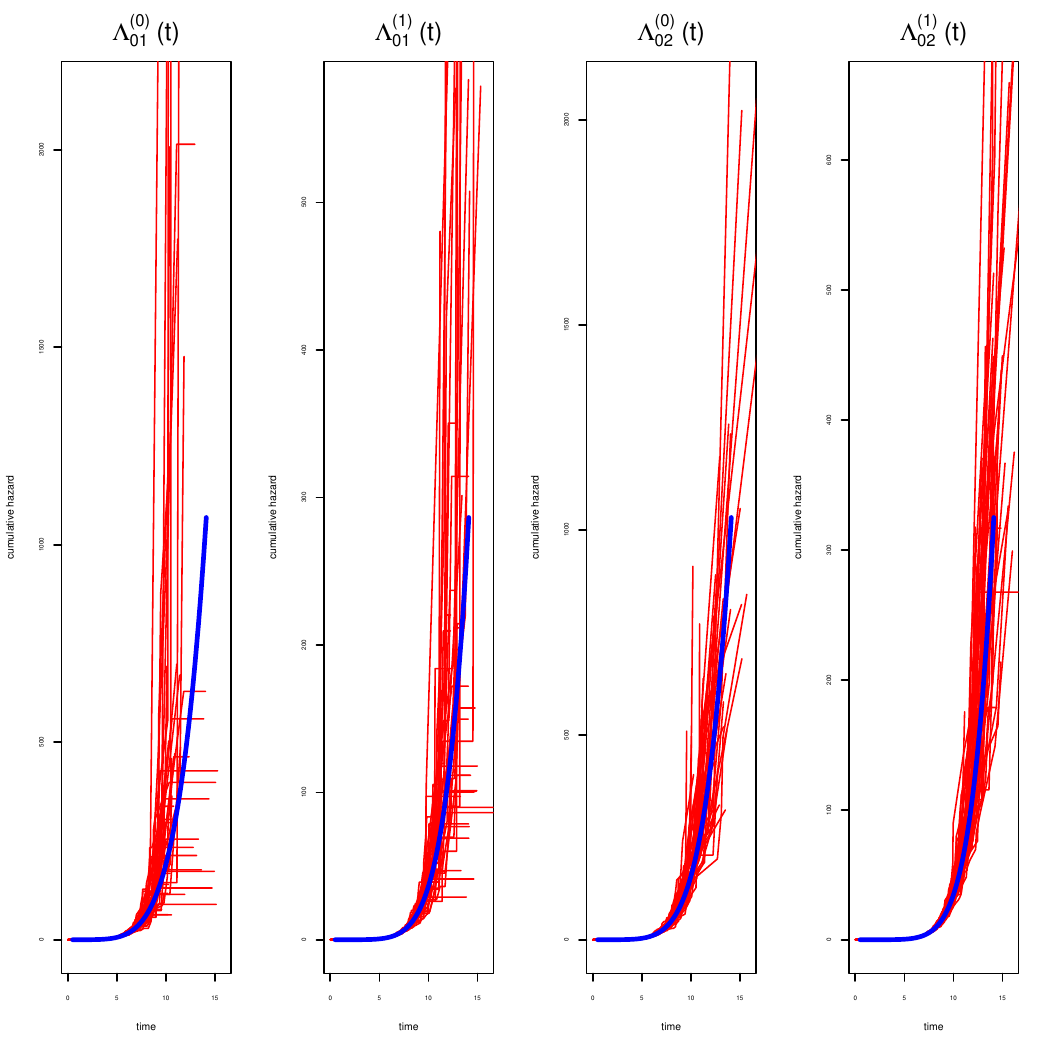}
  \begin{center}
    \texttt{(a) low censoring}
\end{center}

  \medskip
\includegraphics[height=9cm,width=17cm]{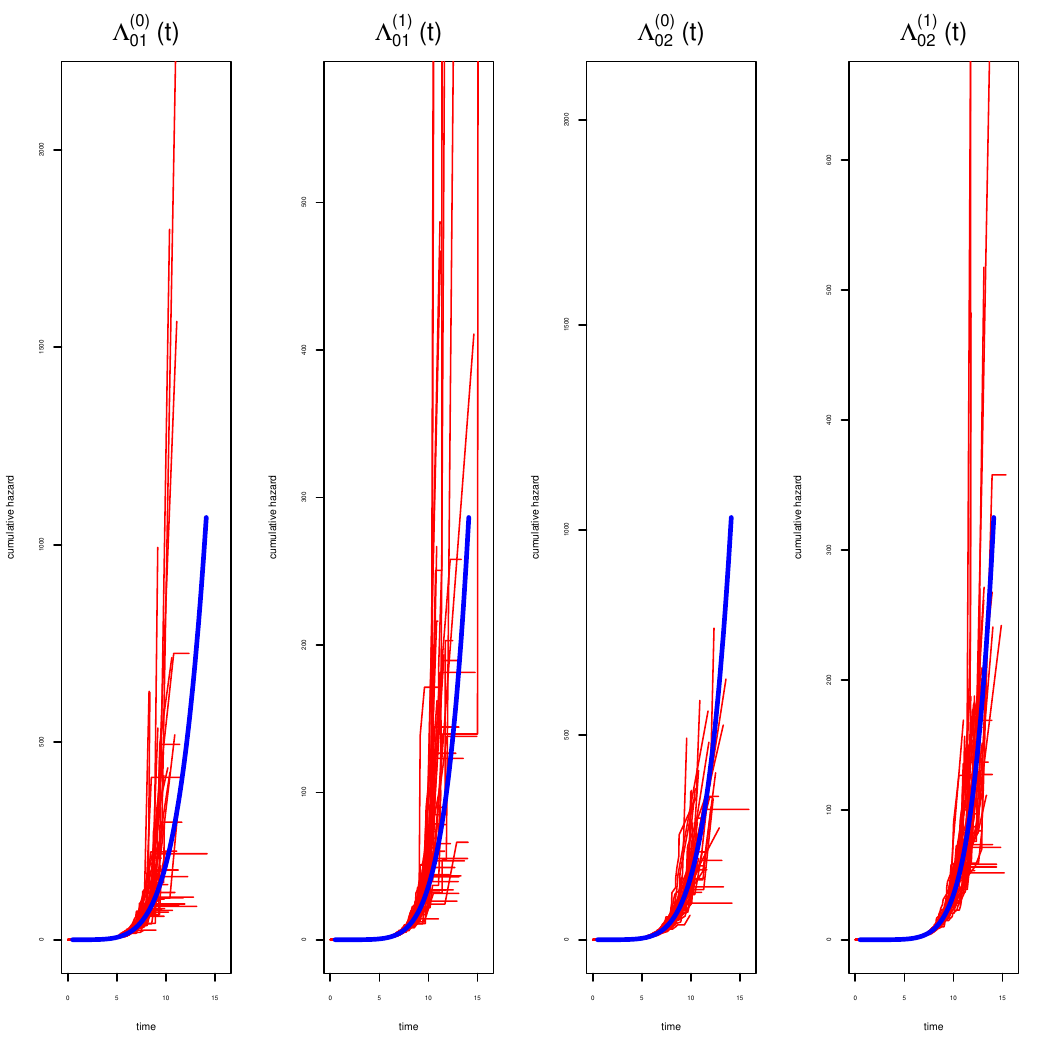}
\begin{center}
    \texttt{(b) high censoring}
\end{center}
\caption{Estimation of baseline cumulative hazard functions for each arm in the first example with $n=1000$ and $\tau=0.3$.}
    \label{fig:bshaz}
\end{figure}

\begin{table}[htbp]
\begin{center}
\caption{Estimation results of regression coefficients and association parameters in  Ex4. The mean biases and Monte Carlo standard derivation (MCSD) of estimated parameters are summarized in this table.}
\label{tab:est_coef_ex4}
\begin{tabular}{lrrrrrrrrrrrr}
\hline\hline
& &\multicolumn{5}{c}{$\tau=0.3$}&&\multicolumn{5}{c}{$\tau=0.6$}\tabularnewline
\cline{3-7}\cline{9-13}
& &\multicolumn{2}{c}{$n=500$}&&\multicolumn{2}{c}{$n=1000$}&&\multicolumn{2}{c}{$n=500$}&&\multicolumn{2}{c}{$n=1000$}\tabularnewline
\cline{3-4}\cline{6-7}\cline{9-10}
\cline{12-13}
 && Bias& MCSD&& Bias& MCSD&& Bias& MCSD&& Bias& MCSD \tabularnewline
\hline
\multicolumn{5}{l}{$\sigma=0.2$}\tabularnewline
$\tau^{(0)}$ &&$ 0.018$&$0.061$&&$ 0.004$&$0.047$&&$-0.031$&$0.044$&&$-0.033$&$0.036$\tabularnewline
$\tau^{(1)}$&&$ 0.014$&$0.057$&&$ 0.008$&$0.044$&&$-0.051$&$0.062$&&$-0.043$&$0.042$\tabularnewline
$\beta_{11}^{(0)}$&&$ 0.004$&$0.241$&&$ 0.027$&$0.156$&&$ 0.018$&$0.302$&&$ 0.032$&$0.202$\tabularnewline
$\beta_{12}^{(0)}$&&$ 0.004$&$0.142$&&$-0.003$&$0.086$&&$ 0.079$&$0.150$&&$ 0.073$&$0.120$\tabularnewline
$\beta_{11}^{(1)}$&&$-0.012$&$0.198$&&$ 0.007$&$0.149$&&$-0.045$&$0.227$&&$-0.051$&$0.149$\tabularnewline
$\beta_{12}^{(1)}$&&$ 0.013$&$0.115$&&$ 0.004$&$0.083$&&$ 0.048$&$0.127$&&$ 0.053$&$0.077$\tabularnewline
$\beta_{21}^{(0)}$&&$ 0.004$&$0.150$&&$ 0.016$&$0.114$&&$ 0.018$&$0.167$&&$ 0.032$&$0.120$\tabularnewline
$\beta_{22}^{(0)}$&&$-0.013$&$0.154$&&$-0.006$&$0.109$&&$ 0.024$&$0.169$&&$ 0.029$&$0.110$\tabularnewline
$\beta_{21}^{(1)}$&&$-0.004$&$0.156$&&$ 0.001$&$0.107$&&$-0.001$&$0.170$&&$ 0.008$&$0.112$\tabularnewline
$\beta_{22}^{(1)}$&&$ 0.007$&$0.114$&&$ 0.001$&$0.073$&&$ 0.045$&$0.126$&&$ 0.043$&$0.080$\tabularnewline
\hline
\multicolumn{5}{l}{$\sigma=0.4$}\tabularnewline
$\tau^{(0)}$ &&$ 0.032$&$0.063$&&$ 0.011$&$0.048$&&$-0.039$&$0.049$&&$-0.039$&$0.039$\tabularnewline
$\tau^{(1)}$ &&$ 0.021$&$0.067$&&$ 0.019$&$0.047$&&$-0.057$&$0.068$&&$-0.050$&$0.059$\tabularnewline
$\beta_{11}^{(0)}$&&$-0.024$&$0.229$&&$-0.019$&$0.166$&&$ 0.008$&$0.291$&&$-0.001$&$0.212$\tabularnewline
$\beta_{12}^{(0)}$&&$-0.005$&$0.130$&&$-0.007$&$0.093$&&$ 0.051$&$0.175$&&$ 0.042$&$0.146$\tabularnewline
$\beta_{11}^{(1)}$&&$-0.010$&$0.204$&&$-0.042$&$0.136$&&$-0.053$&$0.241$&&$-0.070$&$0.179$\tabularnewline
$\beta_{12}^{(1)}$&&$ 0.006$&$0.118$&&$ 0.001$&$0.086$&&$ 0.039$&$0.122$&&$ 0.038$&$0.105$\tabularnewline
$\beta_{21}^{(0)}$&&$-0.001$&$0.174$&&$ 0.005$&$0.118$&&$ 0.009$&$0.176$&&$ 0.012$&$0.114$\tabularnewline
$\beta_{22}^{(0)}$&&$-0.031$&$0.166$&&$-0.021$&$0.119$&&$ 0.006$&$0.178$&&$ 0.005$&$0.129$\tabularnewline
$\beta_{21}^{(1)}$&&$ 0.004$&$0.181$&&$-0.016$&$0.111$&&$-0.006$&$0.201$&&$-0.026$&$0.123$\tabularnewline
$\beta_{22}^{(1)}$&&$-0.002$&$0.121$&&$-0.015$&$0.087$&&$ 0.024$&$0.131$&&$ 0.018$&$0.103$\tabularnewline
\hline
\end{tabular}\end{center}
\end{table}


\begin{figure}[h]
  \medskip
\includegraphics[height=9cm,width=17cm]{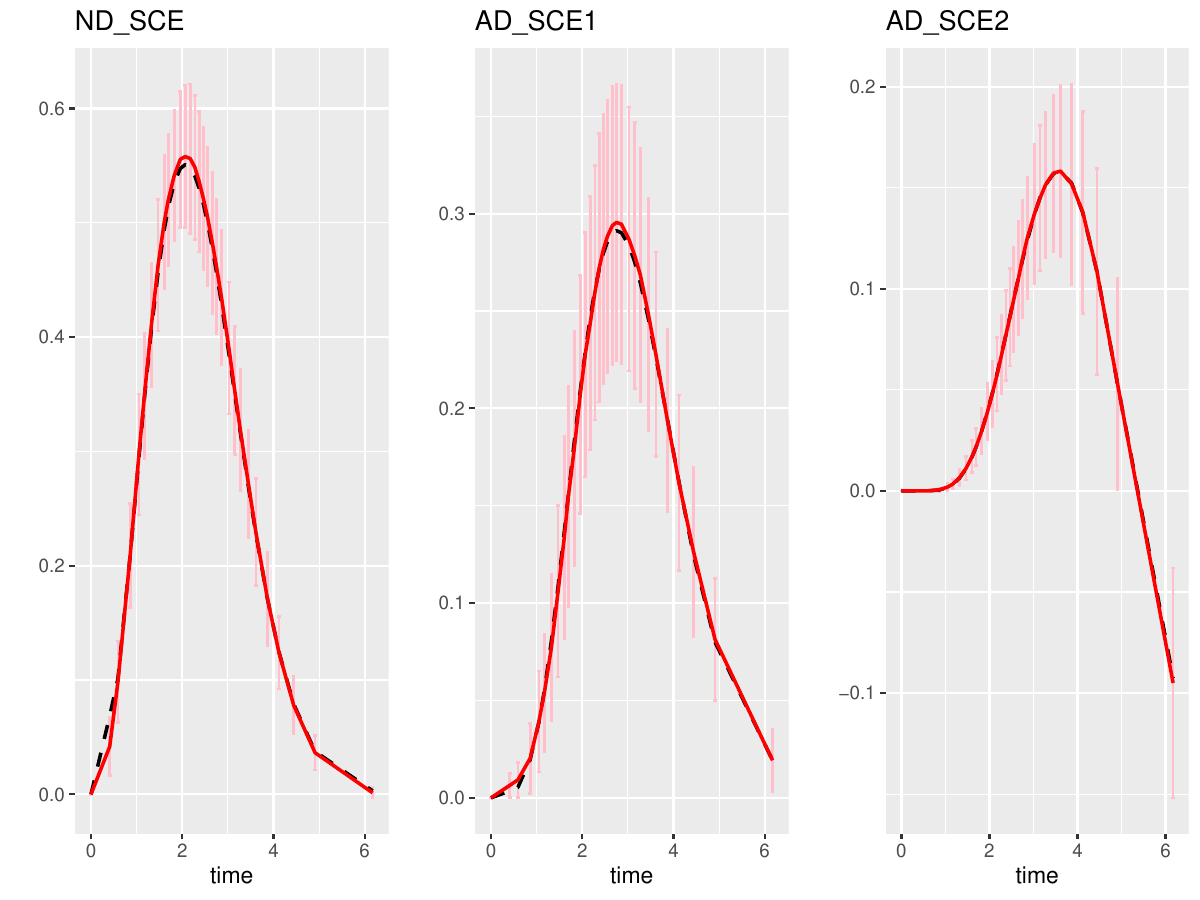}
  \begin{center}
    \texttt{(a) $\tau=0.3$}
\end{center}
  \medskip
\includegraphics[height=9cm,width=17cm]{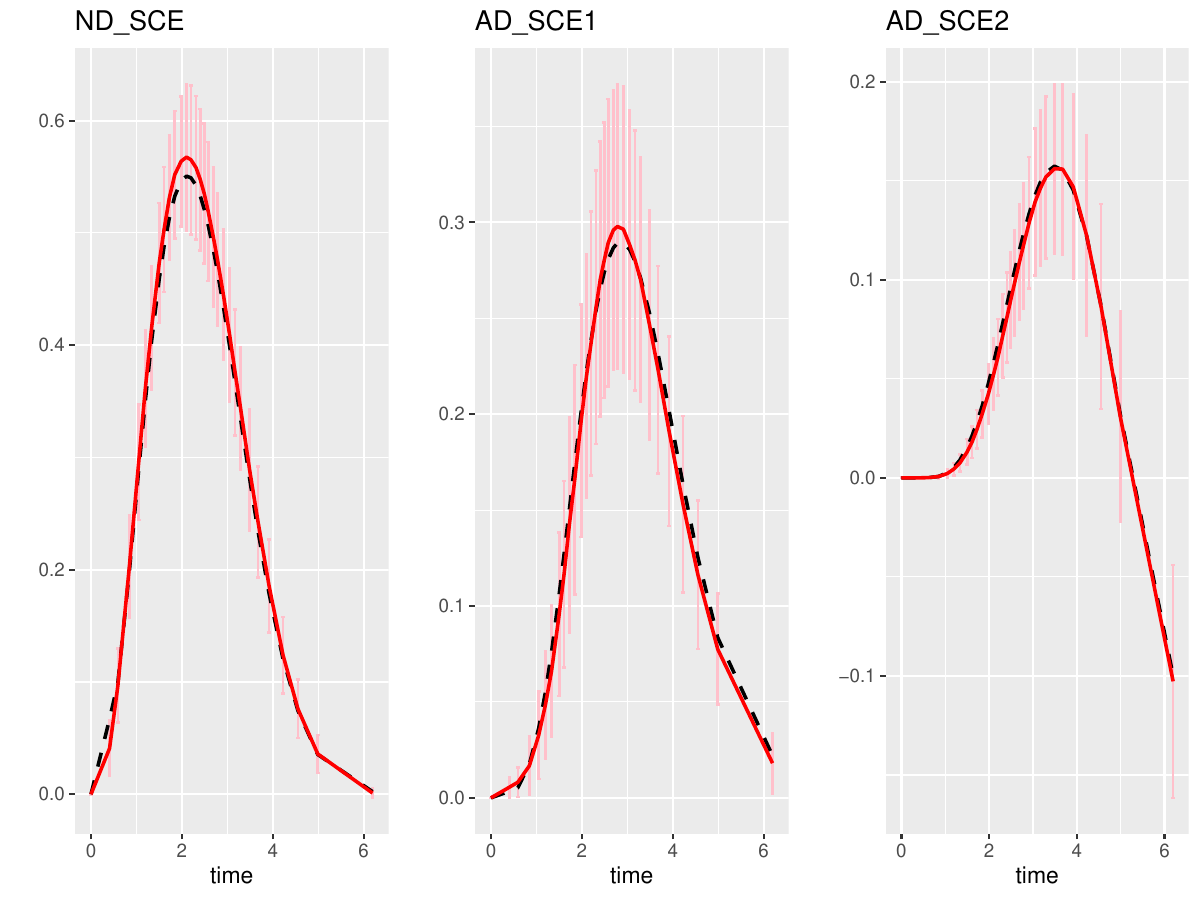}
\begin{center}
    \texttt{(b) $\tau=0.6$}
\end{center}
\caption{Estimation of causal parameters in Ex2 with $\tau=0.3,0.6$ and $n=1000$.}
    \label{fig:causal_ex2}
\end{figure}

\begin{figure}[h]
  \medskip
\includegraphics[height=9cm,width=17cm]{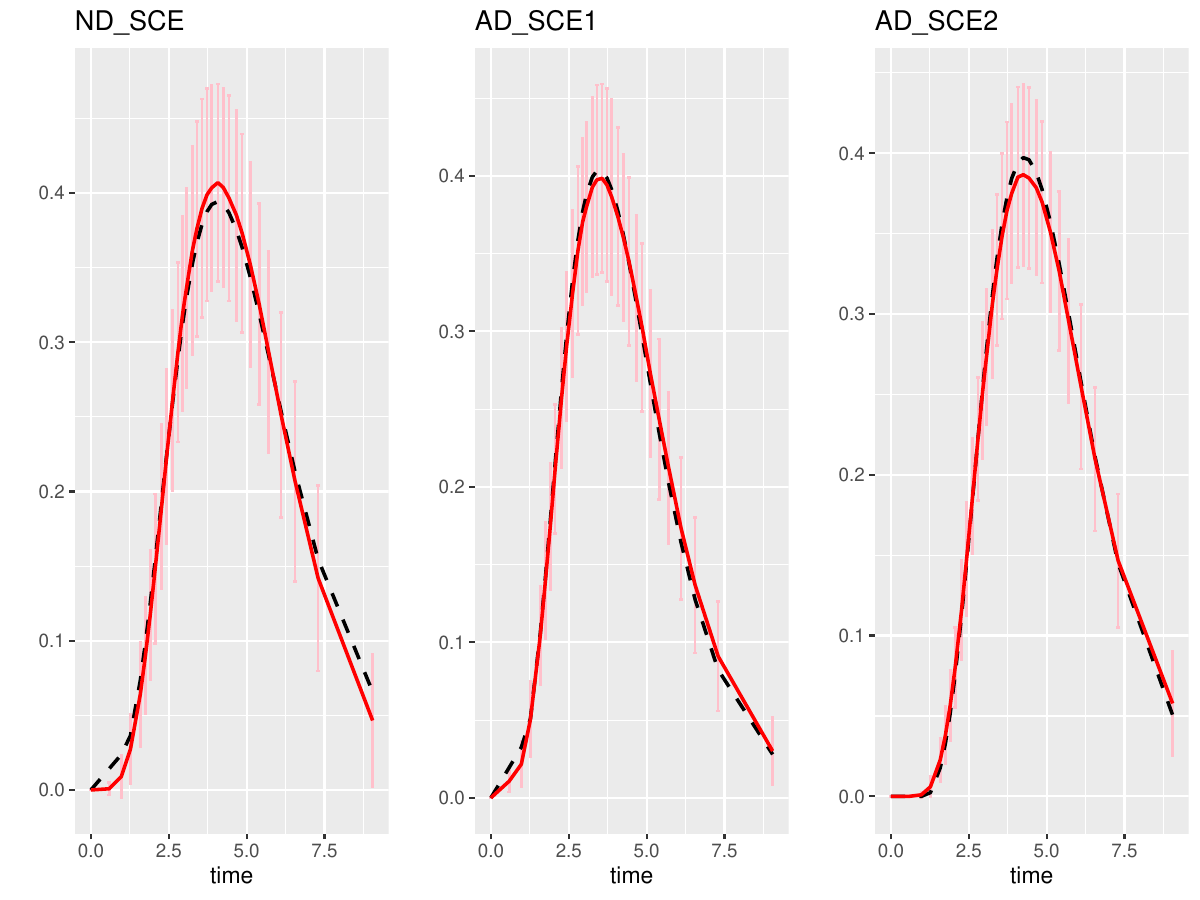}
  \begin{center}
    \texttt{(a) $\tau=0.3$}
\end{center}
  \medskip
\includegraphics[height=9cm,width=17cm]{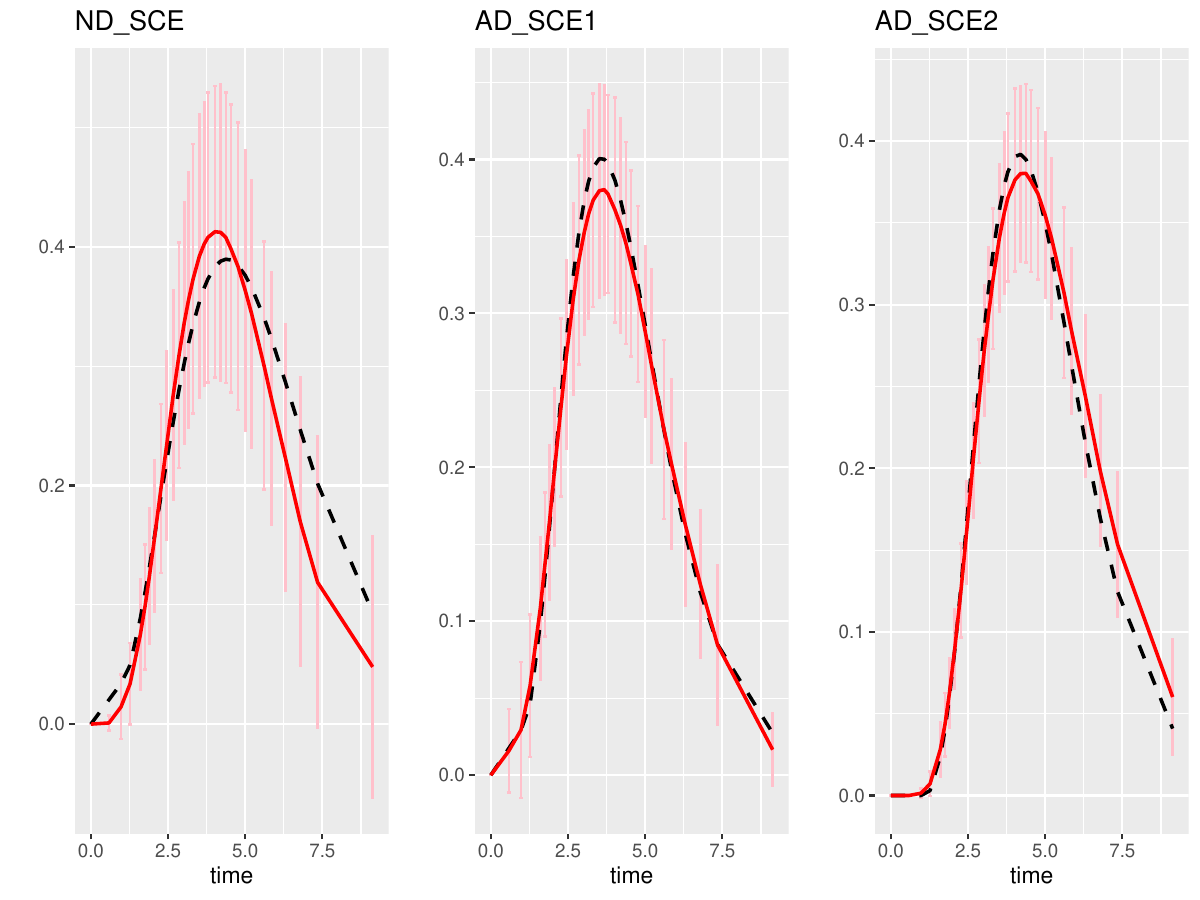}
\begin{center}
    \texttt{(b) $\tau=0.6$}
\end{center}
\caption{Estimation of causal parameters in Ex3 with $\tau=0.3,0.6$ and $n=1000$.}
    \label{fig:causal_ex3}
\end{figure}

\begin{figure}[h]
  \medskip
\includegraphics[height=9cm,width=17cm]{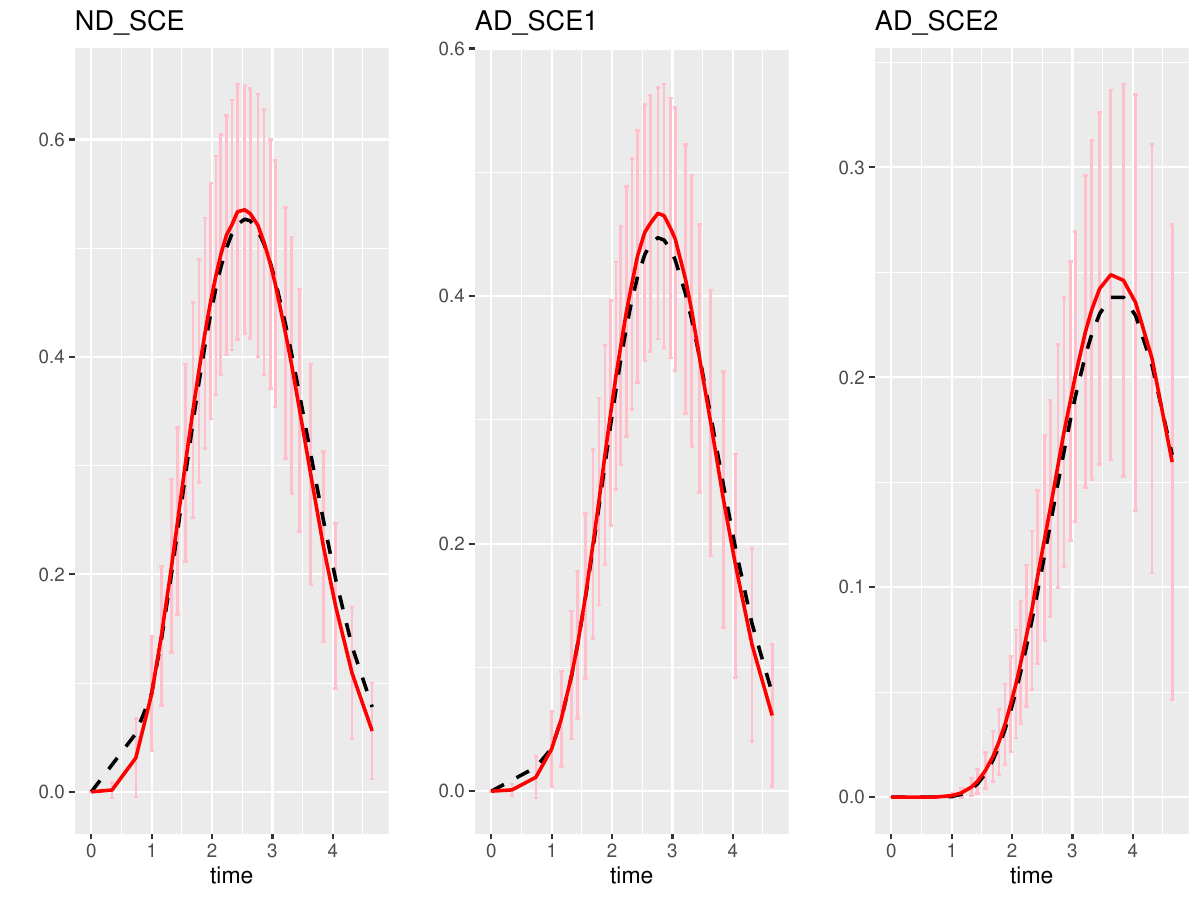}
  \begin{center}
    \texttt{(a) $n=500$}
\end{center}

  \medskip
\includegraphics[height=9cm,width=17cm]{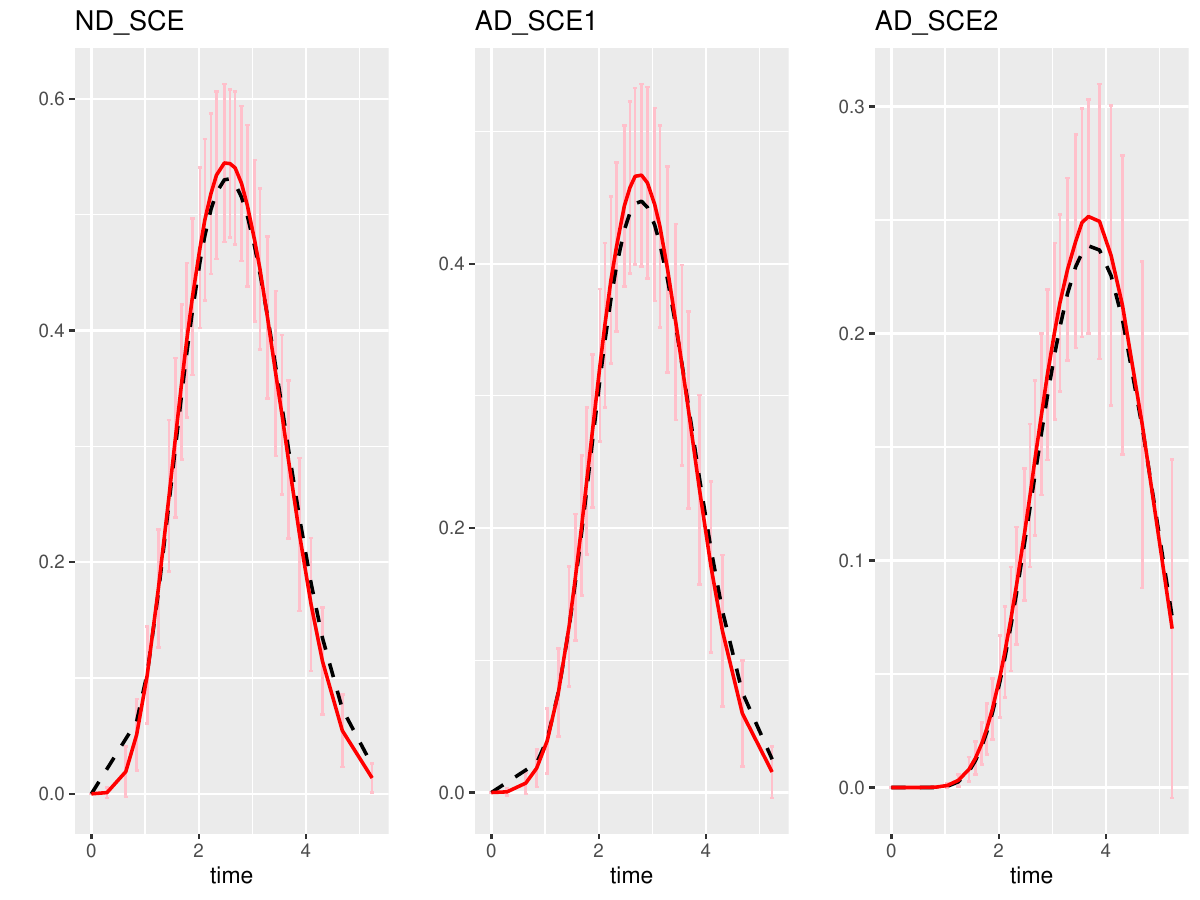}
\begin{center}
    \texttt{(b) $n=1000$}
\end{center}
\caption{Estimation of causal parameters in Ex4 with $\tau=0.3$ and $\sigma=0.2$.}
    \label{fig:causal0.3_ex41}
\end{figure}

\begin{figure}[h]
  \medskip
\includegraphics[height=9cm,width=17cm]{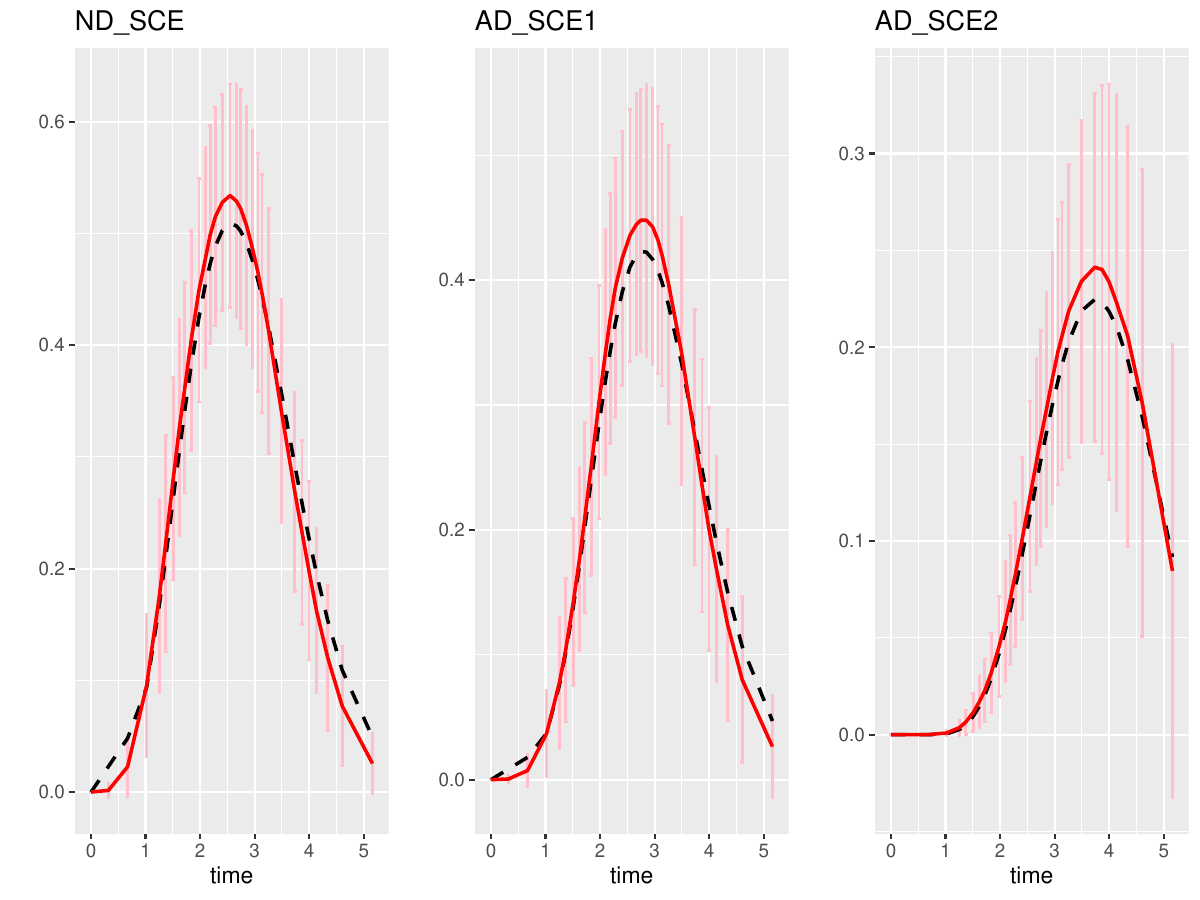}
  \begin{center}
    \texttt{(a) $n=500$}
\end{center}

  \medskip
\includegraphics[height=9cm,width=17cm]{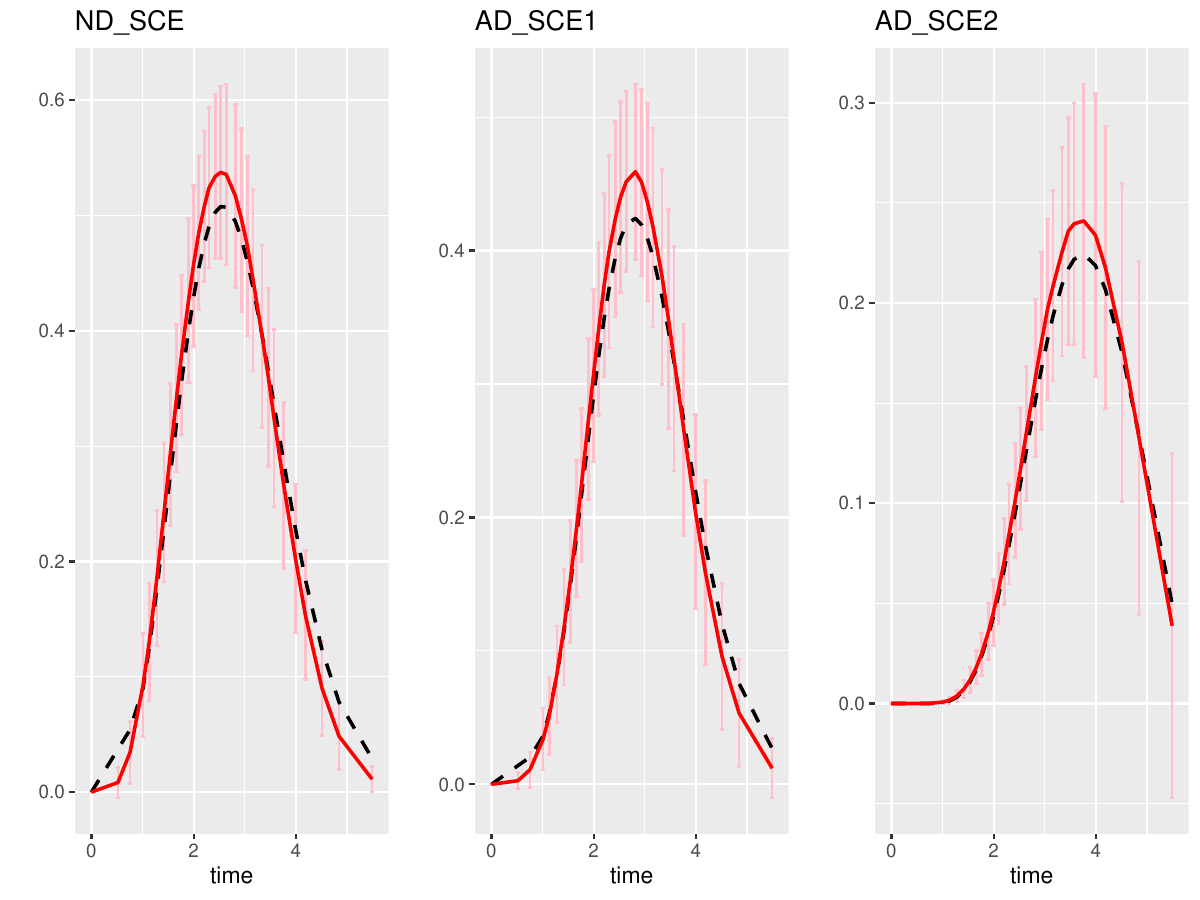}
\begin{center}
    \texttt{(b) $n=1000$}
\end{center}
\caption{Estimation of causal parameters in Ex4 with $\tau=0.3$ and $\sigma=0.4$.}
    \label{fig:causal0.6_ex42}
\end{figure}

\begin{figure}[h]
 \medskip
\includegraphics[height=9cm,width=17cm]{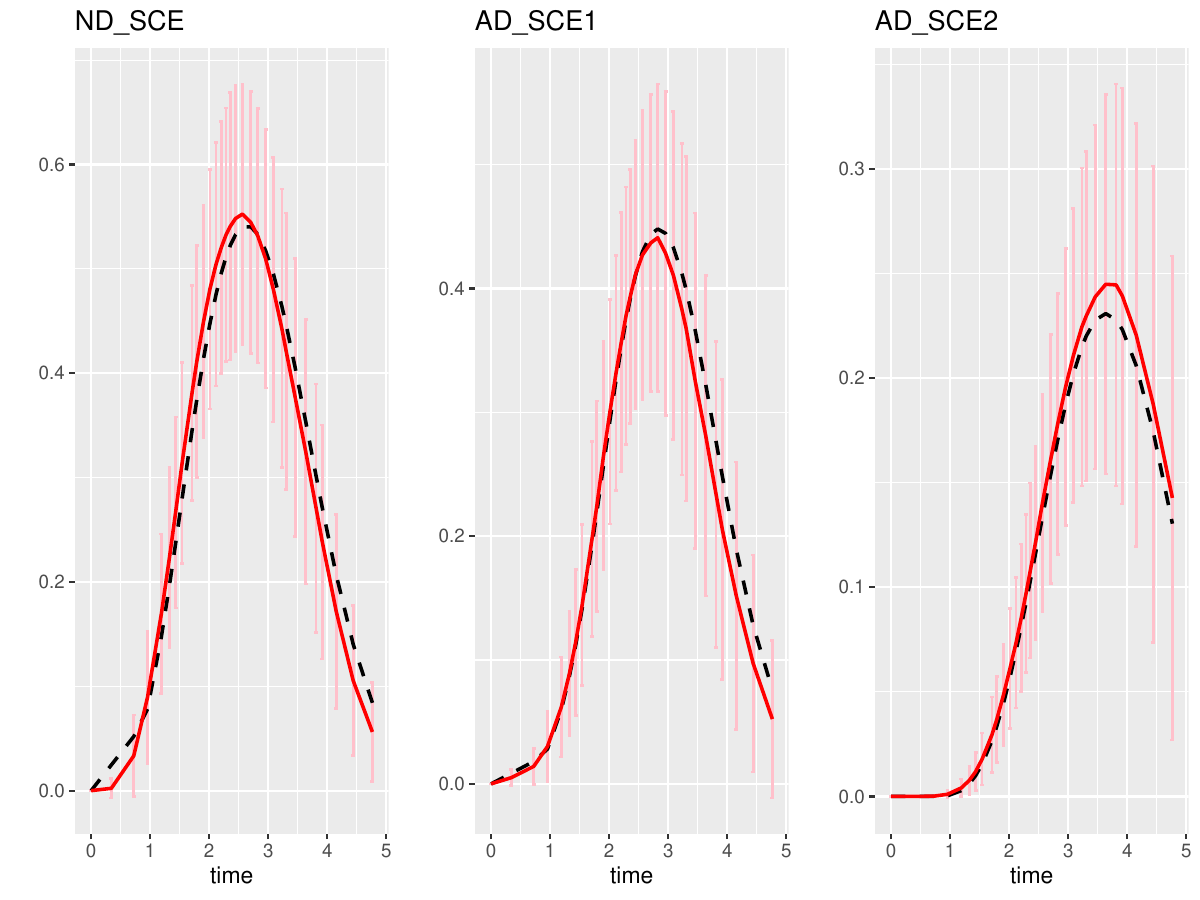}
 \begin{center}
 \texttt{(a) $\sigma=0.2$}
\end{center}

\includegraphics[height=9cm,width=17cm]{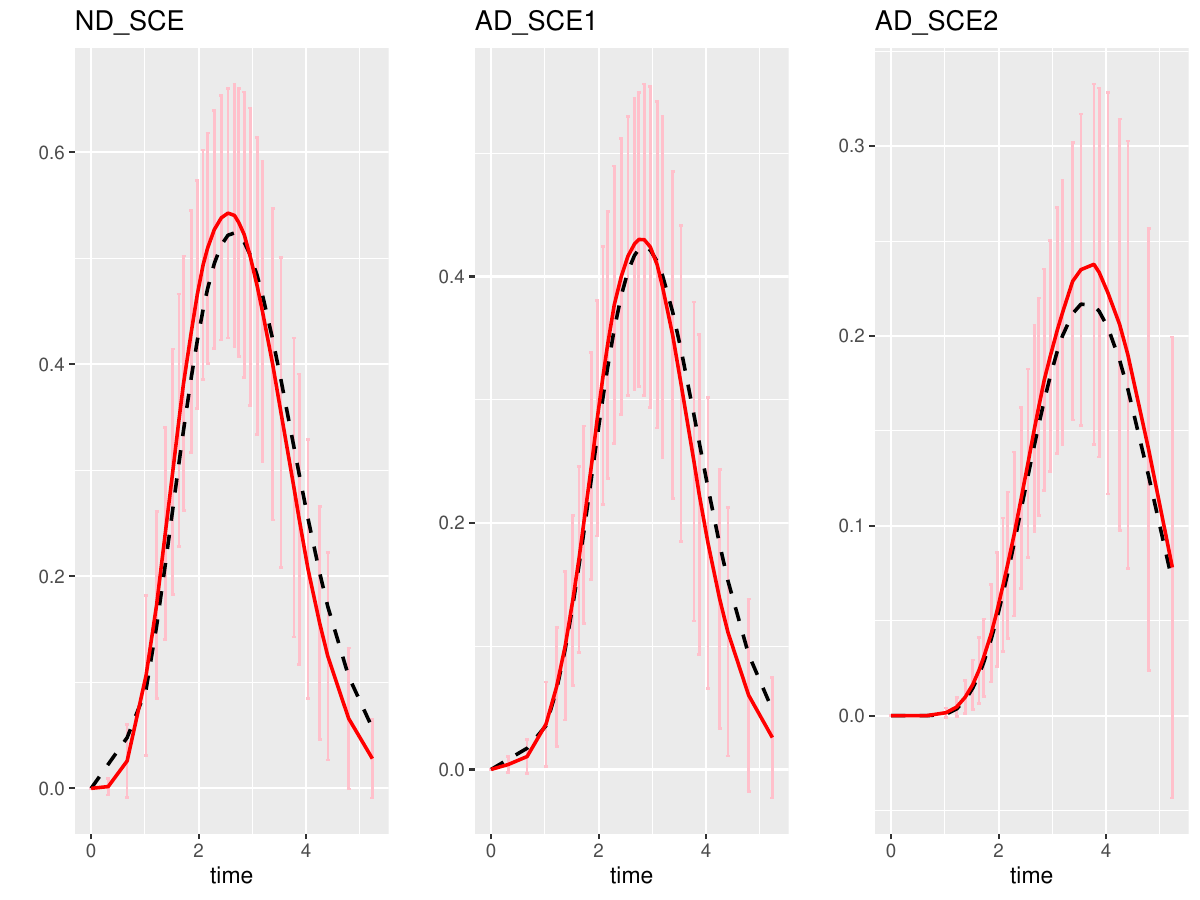}
  \begin{center}
    \texttt{(a) $\sigma=0.4$}
\end{center}

\caption{Estimation of causal parameters in Ex4 with $n=500$, $\tau=0.6$ and $\sigma=0.2/0.4$.}
    \label{fig:causal0.6_ex42}
\end{figure}

\clearpage

\section{Additional results for breast cancer study}
\label{sec:extra real}

\bibliographystyle{biom} 
\bibliography{ref_causal}

\begin{landscape}
\begin{table}[htbp]
\footnotesize
\tabcolsep=5pt
\begin{center}
\caption{Estimation of regression coefficients and Kendall's tau $\tau^{(a)}$ for the breast cancer study under different specified frailty variance $\sigma$, where $a=0$ and 1 represent no treatment and hormone treatment, respectively. }
\label{tab:breast_coef1}
\begin{tabular}{llllllllllllll}
\hline\hline
\multirow{2}*{$\sigma$}&\multirow{2}*{$\tau^{(a)}$}& \multicolumn{6}{c}{RFS$(T_1)$}&\multicolumn{6}{c}{OS$(T_2)$}\tabularnewline
\cmidrule(lr){3-8}\cmidrule(lr){9-14}
&&Age&ER&Meno&Node&NPI&size&Age&ER&Meno&Node&NPI&Size\tabularnewline
\hline
\multicolumn{6}{l}{a=0 (no treatment)}\tabularnewline
$ 0$&$0.574^{**}$&$ 0.005$&$0.457$&$-0.057$&$0.613$&$0.228*$&$0.008$&$0.056^{**}$&$-0.021$&$-0.732^{**}$&$0.909^{**}$&$0.175*$&$0.011$\tabularnewline
$ 0.5$&$0.566^{**}$&$ 0.000$&$0.403$&$ 0.033$&$0.696*$&$0.248*$&$0.006$&$0.055$&$-0.003$&$-0.705^{**}$&$0.981^{**}$&$0.253^{**}$&$0.013$\tabularnewline
$ 1$&$0.438^{**}$&$-0.002$&$0.388$&$-0.025$&$0.762$&$0.319^{**}$&$0.005$&$0.054$&$-0.083$&$-0.730^{*}$&$1.090^{**}$&$0.283^{**}$&$0.012$\tabularnewline
$ 1.5$&$0.548^{**}$&$ 0.000$&$0.334$&$-0.034$&$0.678^{*}$&$0.266^{*}$&$0.008$&$0.054$&$-0.012$&$-0.710^{**}$&$0.927^{**}$&$0.213^{*}$&$0.011$\tabularnewline
$10$&$0.498^{**}$&$-0.005$&$0.369$&$-0.037$&$0.743$&$0.252^{*}$&$0.005$&$0.051$&$-0.011$&$-0.749^{*}$&$0.964^{*}$&$0.197$&$0.011$\tabularnewline
\hline
\multicolumn{6}{l}{a=1 (hormone treatment)}\tabularnewline
$ 0$&$0.675^{**}$&$0.016$&$ 0.008$&$-0.081$&$0.500^{**}$&$0.179^{*}$&$0.021^{**}$&$0.044^{**}$&$-0.071$&$-0.027$&$0.270$&$0.188^{**}$&$0.017^{**}$\tabularnewline
$ 0.5$&$0.655^{**}$&$0.010$&$-0.037$&$-0.134$&$0.515^{**}$&$0.281^{*}$&$0.025^{*}$&$0.045$&$-0.097$&$-0.005$&$0.337^{*}$&$0.236^{**}$&$0.017^{**}$\tabularnewline
$ 1$&$0.505^{**}$&$0.004$&$-0.082$&$-0.124$&$0.559^{*}$&$0.289^{**}$&$0.024^{**}$&$0.042$&$-0.104$&$ 0.093$&$0.312$&$0.239$&$0.019^{**}$\tabularnewline
$ 1.5$&$0.652^{**}$&$0.007$&$-0.082$&$-0.152$&$0.558^{**}$&$0.273^{*}$&$0.025$&$0.044$&$-0.129$&$ 0.008$&$0.355^{*}$&$0.249^{**}$&$0.019^{**}$\tabularnewline
$10$&$0.584^{**}$&$0.004$&$-0.056$&$-0.151$&$0.563^{**}$&$0.270^{**}$&$0.024$&$0.041$&$-0.111$&$ 0.000$&$0.329^{*}$&$0.226^{**}$&$0.016^{**}$\tabularnewline
\hline
\multicolumn{14}{l}{* and ** indicate significance computed by 100  Bootstrap replicates at levels 0.1 and 0.05, respectively.}\tabularnewline
\end{tabular}\end{center}
\end{table}
\end{landscape}


\begin{landscape}
\begin{table}[htbp]
\footnotesize
\tabcolsep=5pt
\begin{center}
\caption{Estimation of regression coefficients and Kendall's tau $\tau^{(a)}$ for the breast cancer study under different specified frailty variance $\sigma$, where $a=0$ and 1 represent hormone treatment and combined treatment of hormone and radio, respectively. }
\label{tab:breast_coef}
\begin{tabular}{llllllllllllllll}
\hline\hline
\multirow{2}*{$\sigma$}&\multirow{2}*{$\tau^{(a)}$}& \multicolumn{7}{c}{RFS$(T_1)$}&\multicolumn{7}{c}{OS$(T_2)$}\tabularnewline
\cmidrule(lr){3-9}\cmidrule(lr){10-16}
&&{Age}&{PR}&{HER2}&{Meno}&{Node}&{NPI}&{Size}&{Age}&{PR}&{HER2}&{Meno}&{Node}&{NPI}&{Size}\tabularnewline
\hline
\multicolumn{10}{l}{a=0 (hormone treatment only)}\tabularnewline
0&0.676**&0.017*&0.005&0.262&-0.136&0.49**&0.182*&0.022**&0.044**&0.081&0.15&-0.015&0.256&0.194**&0.017**\tabularnewline
0.5&0.674**&0.01&-0.042&0.243&-0.18&0.536**&0.277**&0.023&0.045&-0.042&0.198&0.004&0.344*&0.237**&0.02**\tabularnewline
1&0.553**&0.008&-0.031&0.252&-0.188&0.552**&0.282**&0.028**&0.042&-0.052&0.186&0.018&0.293&0.251**&0.021**\tabularnewline
1.5&0.62**&0.008&-0.067&0.23&-0.174&0.588**&0.277&0.027**&0.044&0.042&0.194&-0.097&0.335&0.22**&0.018**\tabularnewline
10&0.585**&0.005&-0.079&0.23&-0.189&0.618**&0.261*&0.024**&0.04&0.068&0.177&-0.021&0.279&0.221**&0.016\tabularnewline
\hline
\multicolumn{10}{l}{a=1 (combined treatment of hormone and radio)}\tabularnewline
0&0.65**&0.011&-0.461**&0.709**&-1.002**&0.468&0.08&0.015**&0.054**&-0.203&0.536&-0.499&0.336&0.103&0.014**\tabularnewline
0.5&0.637**&0.001&-0.377&0.515&-0.935&0.468&0.187&0.015&0.054&-0.225&0.604&-0.448&0.414&0.122&0.015*\tabularnewline
1&0.554**&0.002&-0.44&0.797**&-0.942&0.365&0.199&0.015*&0.054&-0.21&0.65*&-0.509&0.383&0.137&0.012\tabularnewline
1.5&0.598**&0.001&-0.446&0.719**&-0.942&0.452&0.195&0.016&0.053&-0.215&0.625**&-0.493&0.375&0.104&0.012\tabularnewline
10&0.435**&0.099&-0.376&0.808**&-1.058*&0.367&0.23&0.019*&0.054&-0.279&0.576*&-0.404&0.359&0.073&0.016\tabularnewline
\hline
\multicolumn{14}{l}{* and ** indicate significance computed by 100  Bootstrap replicates at levels 0.1 and 0.05, respectively.}\tabularnewline
\end{tabular}\end{center}
\end{table}
\end{landscape}

\begin{landscape}
\begin{figure}
    \centering
    \includegraphics[height=15cm,width=28cm]{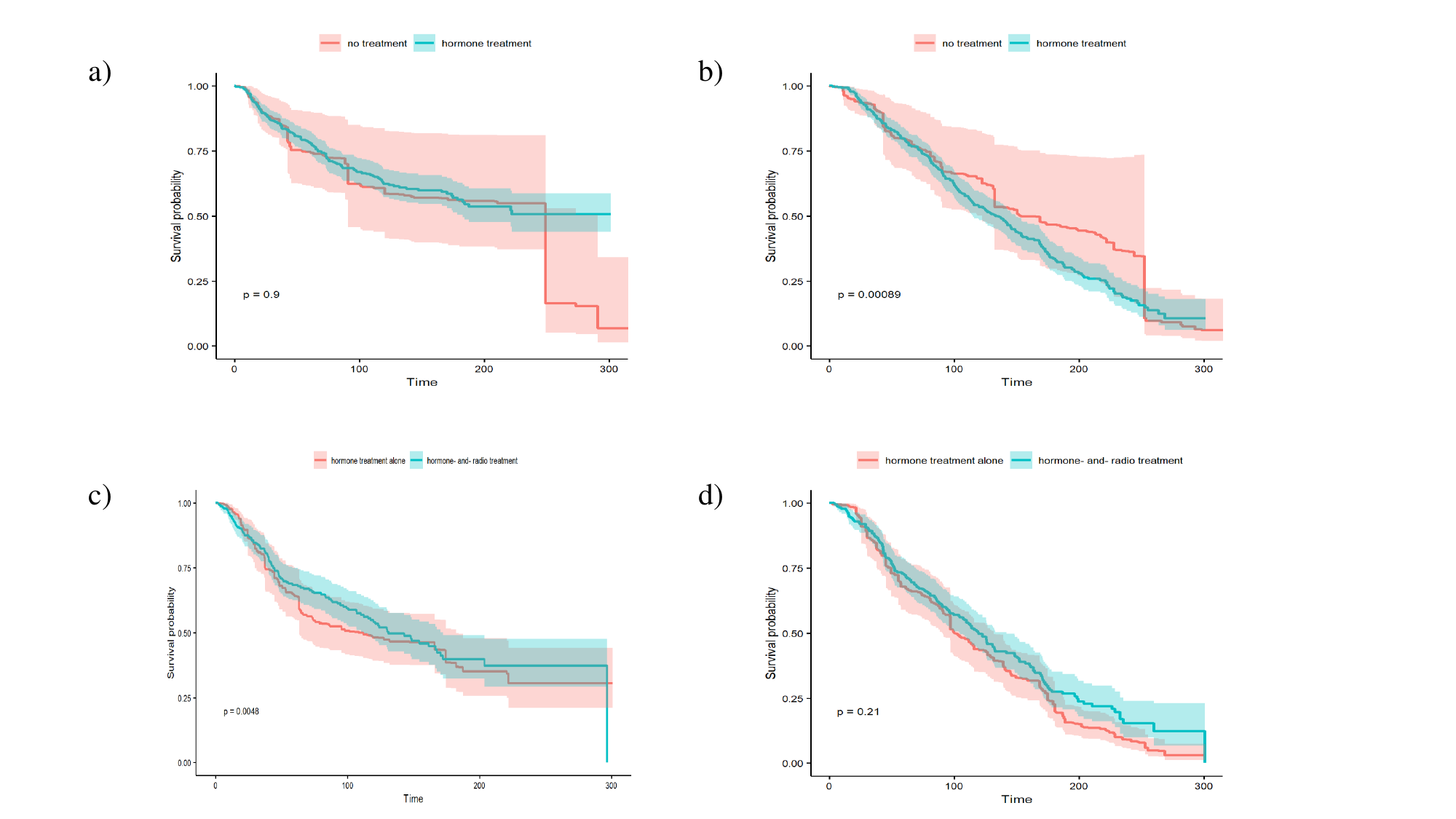}
    \caption{Kaplan–Meier survival curves obtained using propensity score matching method. a) and c) are KM survival curves of RFS by ignoring dependent censoring. b) and d) are  KM survival curves of OS for breast cancer patients. }
    \label{fig:matchkm}
\end{figure}
\end{landscape}